\documentclass[suppldata]{interact}

\renewcommand{\titlefont}{%
  \fontsize{16}{19}\selectfont
  \bfseries
  \raggedright
  \boldmath
}

\usepackage{epstopdf}
\usepackage[caption=false]{subfig}

\usepackage[longnamesfirst,sort]{natbib}
\bibpunct[, ]{(}{)}{;}{a}{,}{,}
\renewcommand\bibfont{\fontsize{10}{12}\selectfont}

\renewcommand{\sectionfont}{%
  \fontsize{15}{18}\selectfont
  \bfseries
  \raggedright
  \boldmath
}

\usepackage[most]{tcolorbox}
\usepackage{natbib}
\usepackage{etoolbox}

\AtBeginEnvironment{thebibliography}{%
  \interlinepenalty=10000
}
\usepackage{graphicx}
\usepackage{moreverb,url}
\usepackage{xcolor}
\usepackage{float}

\theoremstyle{plain}

\theoremstyle{definition}

\theoremstyle{remark}

\usepackage{musicography} 

\newcommand{\citeauthorpos}[1]{{\citeauthor{#1}'s}}

\newcommand{\citepos}[1]{{\citeauthorpos{#1}~\citeyearpar{#1}}}

\usepackage{etoolbox}

\makeatletter
\patchcmd{\@maketitle}
  {\vspace*{36pt}}
  {\vspace*{0pt}}
  {}{}
\makeatother

\begin{document}


\title{The evolution of inharmonicity and noisiness in contemporary popular music}

\author{
\name{Emmanuel Deruty\textsuperscript{a,b}\thanks{CONTACT Emmanuel Deruty. Email: emmanuel.deruty@sony.com}, David Meredith\textsuperscript{b}, and Stefan Lattner\textsuperscript{a}}
\affil{\textsuperscript{a}Sony Computer Science Laboratories, 6 rue Amyot, 75005 Paris, France\\\textsuperscript{b}Department of Architecture, Design and Media Technology, Aalborg University, Rendsburggade 14, 9000 Aalborg, Denmark}
}

\maketitle

\vspace{1cm}

\begin{tcolorbox}[
  colback=gray!10,
  colframe=gray!60,
  boxrule=0.5pt,
  arc=1mm
]
\noindent \textbf{Published version:} Deruty, E., Meredith, D., \& Lattner, S. (2023). The evolution of inharmonicity and noisiness in contemporary popular music. \textit{Journal of New Music Research}, 52(5), 382–409. \\\url{https://doi.org/10.1080/09298215.2024.2434461}
\end{tcolorbox}

\vspace{1cm}


\begin{abstract}

Much of Western classical music relies on instruments based on acoustic resonance, which produce harmonic or quasi-harmonic sounds. In contrast, since the mid-twentieth century, popular music has increasingly been produced in recording studios, where it is not bound by the constraints of harmonic sounds. In this study, we use modified MPEG-7 features to explore and characterise the evolution of noise and inharmonicity in popular music since 1961. We place this evolution in the context of other broad categories of music, including Western classical piano music, orchestral music, and \textit{musique concr\`ete}. We introduce new features that distinguish between inharmonicity caused by noise and that resulting from interactions between discrete partials. Our analysis reveals that the history of popular music since 1961 can be divided into three phases. From 1961 to 1972, inharmonicity in popular music, initially only slightly higher than in orchestral music, increased significantly. Between 1972 and 1986, this rise in inharmonicity was accompanied by an increase in noise, but since 1986, both inharmonicity and noise have moderately decreased. In recent years (up to 2020), popular music has remained much more inharmonic than popular music from the 1960s or orchestral music involving acoustic resonance instruments. However, it has become less noisy, with noise levels comparable to those of orchestral music. We relate these trends to the evolution of music production techniques. In particular, the use of multi-tracking may explain the higher inharmonicity in popular music compared to orchestral music. We illustrate these trends with analyses of key artists and tracks.

\end{abstract}




 \newpage

\section{Introduction}\label{sec:introduction}

Most musical instruments that perform Western classical music are based on the acoustic resonance principle. Such instruments produce harmonic or quasi-harmonic complex tones. In contemporary popular music, the means of production are more diversified. Drums play a key role, and they do not typically produce quasi-harmonic complex tones. Instruments may be based on electronic or digital sound production techniques. The instruments' outputs may be heavily processed with effects. Musicians may not even use instruments and produce the entirety of the music on a computer. 
As a result, it may not be possible to construct an efficient and appropriate description of the signal corresponding to a contemporary popular music track in terms of harmonic or quasi-harmonic complex tones.

In the spectral domain, a quasi-harmonic complex tone may be described as a series of partials whose frequencies are close to being integer multiples of one particular frequency (the fundamental). As observed by \citet[pp.~22,~24]{lavengood2017new}, there are at least two ways audio content may diverge from being quasi-harmonic: (1) the energy may not be centered around partials, in which case the signal may be described as `noisy'; or (2) the frequencies of the partials may not be harmonically related, in which case the tone may be described as `inharmonic'.

In this paper, we trace the evolution over time of the relative amounts of inharmonicity and noise in various audio datasets. Our main focus is contemporary popular music (CPM), but we also compare this class of music with Western classical piano music, Western orchestral music, and \textit{music concr\`ete}.

One key conclusion of this study is that contemporary popular music is more inharmonic than the music from the three other datasets considered. Our diachronic analysis of the evolution of noisiness and inharmonicity in CPM leads to the conclusion that, between 1961 and 1986, recorded music evolved in a general direction that goes from acoustic tonal instruments playing content from a musical score (or, at least, that can be transcribed to a score), to heavily processed, noisy, and inharmonic content made in the recording studio and including drums. In more recent decades, music has become slightly more harmonic.

The remainder of this paper is structured as follows. Section~\ref{sec:backgroundandmotivation} introduces the motivation for examining noise and inharmonicity in contemporary popular music, and elaborates on the relationship between the means of production and the resulting music. Section~\ref{sec:datasetsfeaturesanalysis} forms the core of the article. It describes the datasets (Section~\ref{subsec:datasets}) and the signal features used in the study (Section~\ref{subsec:features}), before presenting a corpus-based study of inharmonicity and noisiness in CPM (Section~\ref{sec:results}). This section is complemented by references to other diachronic studies and remarks on the mastering process. Section~\ref{sec:confrontation} provides detailed analyses of specific music tracks from the perspective proposed in the core sections. 
The paper concludes by summarising the findings from all sections.
The study is complemented by two appendices. 
Appendix~\ref{sec:appendix1} details the signal characteristics that influence the inharmonic character of music and discusses the link between inharmonicity and acoustic beating (or acoustical `grit'). It then substantiates the definitions of inharmonicity and `noisiness' used in the core sections. Appendix~\ref{sec:weighting} explains the influence of the psychoacoustic weighting used in the core sections.


\section{Background and motivation}\label{sec:backgroundandmotivation}

\subsection{On pitch and (in)harmonicity}\label{sec:pitchandinharm}

A defining property of a harmonic tone is that it has a fundamental frequency or $f_0$. A classic music information retrieval task consists of estimating the fundamental frequency of a sound, a task often considered to be synonymous with `pitch tracking' or `pitch estimation' \citep{kim2018crepe,riou2023pesto}. The models performing such tasks are trained on harmonic or quasi-harmonic audio. For instance, \citet{riou2023pesto} uses MIR-1K, a dataset of quasi-harmonic samples \citep{hsu2009improvement} and MDB-stem-synth, a dataset of `exact multiples of the $f_0$'  \citep{salamon2017analysis}.

Another classic music information retrieval task is audio-to-MIDI alignment \citep{raffel2016optimizing}, in which MIDI files are aligned with the corresponding audio content.
 MIDI files have been referred to as symbolic `versions' of pieces of music \citep{ewert2012towards,raffel2016optimizing}, or as `transcription[s]' \citep{raffel2016learning,turetsky2003ground,benetos2018automatic} of the pitched content of the music. Such content has been characterised as `harmonic' \citep{ono2008real} and likely to be played with `harmonic instruments' \citep{ewert2014score}.

\begin{figure}[htbp]
  \centering
  \includegraphics[width=1\columnwidth]{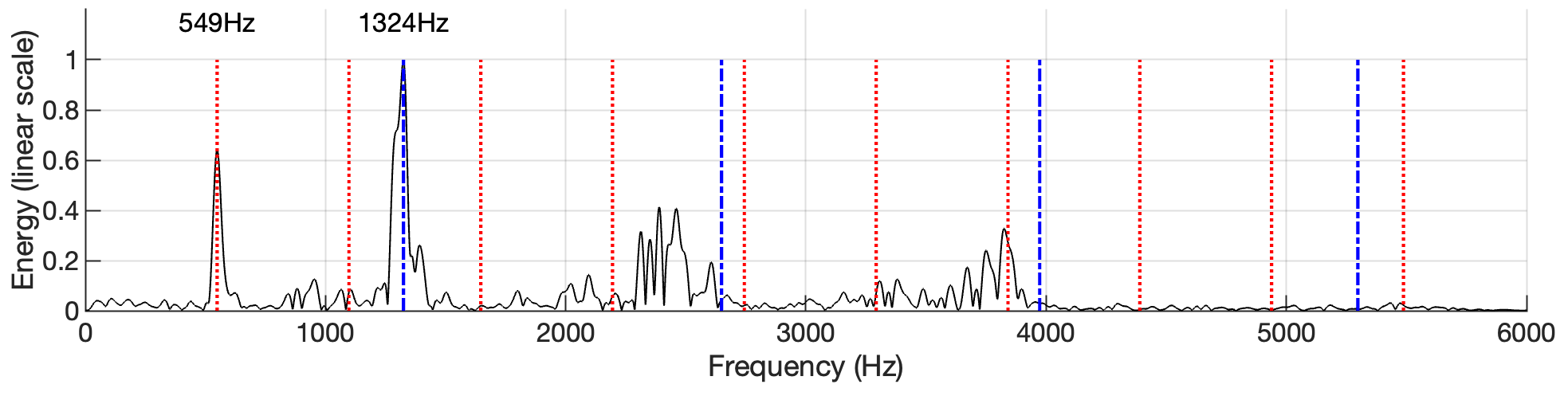}
  \caption{Power spectrum for the first bass `note' in Primaal's `<Fire!'. The vertical red and blue lines represent the fundamental and expected positions of the harmonics for the two highest spectral peaks.}
\label{fig:fire}
\end{figure}

In the pitch tracking and audio-to-midi alignment tasks, pitched content is generally assumed to be harmonic or quasi-harmonic. However, closer examination of the actual practice of successful professional music producers suggests that this assumption does not hold in general---at least for contemporary popular music. For example, Figure~\ref{fig:fire} shows the power spectrum of one bass `note' in the 2023 track `¡Fire!' by Primaal \citep{primaal2023fire}. We write `note' in scare quotes, as its frequency content is not at all harmonic, even though it provides a strong impression of pitch. This example is not unique---there are many such examples of sounds in Primaal's productions that evoke a sense of pitch, while being highly inharmonic. Primaal is a brand of the Hyper Music production company \citep{hypermusic} mentioned by  \citet{deruty2022development}. In 2022 and 2023, Primaal authored music for commercials commissioned by brands such as L'Or\'eal, Adidas, Vichy,  Honda, GoPro and Chanel. The prominence of these brands and the great size and diversity  of the intended audience for this music suggests that Primaal's productions might reasonably be considered to represent important current trends in the music industry.

\vspace{1cm}

\newpage

There therefore seems to be a discrepancy between such examples of the use of inharmonicity in real-world music production and the standard view within music information retrieval that pitched content corresponds to harmonic tones. This discrepancy provides motivation for exploring the extent to which pitch-evoking sounds in CPM are harmonic.


\subsection{On the influence of technology on music}\label{sec:justification}

The example in Figure~\ref{fig:fire} illustrates how some of the resulting sounds in CPM can only be achieved using specific technologies. The example features audio produced by a composite setup of instruments: a Spectrasonics Omnisphere bass superimposed on a Eurorack setup using FM synthesis, resonators, and ring modulators. Producing this with acoustically resonant instruments would have been challenging, if not practically impossible.

This section explores the broader relationship between technology and music. In Section~\ref{subsec:resources}, we discuss the link between performance resources and musical output. Section~\ref{subsec:utopian} briefly covers modern music production technology. In Section~\ref{subsec:extent}, we explain why we have focused on inharmonicity and noise, particularly as CPM may be moving away from using musical notes as primary structuring elements.

\subsubsection{Resources of the performance and musical idioms}\label{subsec:resources}

Music involves what \citet{parry1911style} calls the \textit{resources of the performance}, such as the human voice and musical instruments. In Western classical music, many instruments are based on acoustic resonators, like vibrating strings or air columns, which produce waveforms with strong periodicity. These waveforms are often modelled as the sum of components based on a fundamental frequency and its overtones, which are (close to) integer multiples of that fundamental \citep{young1952inharmonicity}.

According to \citet[p.~103]{huron2009characterizing}, \textit{idiomaticism} is, `of all the ways a given musical goal or effect may be achieved, the method employed by the composer/musician [that] is one of the least difficult'. For instruments with acoustic resonators, producing harmonic or quasi-harmonic sounds is easy, making these sounds \textit{idiomatic} to such instruments. 

Such sounds have formed the basis of much musical discourse. In acoustic resonators, pitch impressions from harmonic or quasi-harmonic sounds are perceived as musical notes, which are used to construct phrases and chords. According to \citepos{pascall2001style} terminology, the idiom lies in the musical phrases and chords enabled by the instrument. The properties of resonator-produced sounds can shape musical discourse. For example, harmonic positioning influences musical temperament \citep{barbour2004tuning}, and Western counterpoint rules, such as avoiding parallel unisons, octaves, and fifths, relate to the waveform properties of resonators \citep{huron2001tone}.

Technological progress expands musical idioms by providing more resources. In the 20\textsuperscript{th} century, signal analysis allowed musicians like Gérard Grisey to create \textit{spectral music}, where scores derive from harmonic positions of acoustic instruments, with deviations creating a gradation from harmonic to inharmonic textures \citep{rose1996introduction}. Here, the idiom in \citet{huron2009characterizing}'s sense is orchestral texture, while in \citet{pascall2001style}'s sense, it is gradation.

The 20\textsuperscript{th} and 21\textsuperscript{st} centuries saw major innovations in music technology \citep{webster2002historical}, such as the Ondes Martenot (1928) \citep{orton2001martenot}, the Moog synthesiser (1964) \citep{porcaro2001moog}, and the FM-based Yamaha DX7 (1983) \citep{mattis2001chowning}, pivotal in '80s popular music \citep[p. 36]{lavengood2017new}. Sound recording technology, evolving since 1877 \citep{mumma2001recording}, also impacted musical idioms. For instance, continuously shifting frequencies are idiomatic to the Ondes Martenot \citep{huron2009characterizing}, while distortion, linked to analog circuitry, defines genres like heavy metal.

\subsubsection{The `utopian vision of a boundless space for musical exploration'}\label{subsec:utopian}

Technological innovation affects more than just the advent of specific idioms. Music-related technology takes various forms and influences different stages of music production and engagement. One such area is the compositional process. For example, before the 16\textsuperscript{th} century, paper was expensive, and erasable pencils did not exist \citep{charlton2001score}, limiting composers' ability to draft and sketch musical ideas \citep{bent2001notation}. The introduction of cheaper paper and pencils transformed composition.

More recently, technological progress has allowed the recording studio to become a creative tool in its own right \citep{moylan2014understanding}. \citet{spicer2004ac} links `accumulative' and `cumulative' forms in pop/rock music to advances in recording technology. The 1990s saw the rise of home studios, making professional practices accessible. Almost anyone with a computer could assume the role of a musician--engineer hybrid \citep{pras2013impact,bell2014trial}, with a focus on micro-manipulating digital audio \citep{theberge2001plugged}.

We argued that using musical notes is idiomatic to acoustic resonators, which impose the \textit{constraint} \citep{mcpherson2020idiomatic} that music involves `notes'—symbols representing harmonic or quasi-harmonic tones based on a fundamental frequency that stays constant during the note's duration. In the studio, this constraint vanishes. A key question is what idiom musicians use without this constraint: `even if we were to achieve the utopian vision of a boundless space for musical exploration, we would still be left with the question of what possibilities musicians would choose to explore within it' \citep{mcpherson2020idiomatic}.

\subsubsection{To what extent do idioms of popular music involve musical notes?}\label{subsec:extent}

\textit{Musique concr\`ete} exemplifies music where musicians have abandoned the musical note as a building block, working in the recording studio \citep{schaeffer2020musique}. This genre, derived from \citepos{schaeffer1966traite} experiments, often lacks clear pitch sensations, yet it is not devoid of pitch. As \citet{yost2009pitch} states, `[m]usic without pitch would be drumbeats, speech without pitch processing would be whispers'. This does not entirely apply to \textit{musique concr\`ete}, as seen in works like Pierre Henry's {\em Temps de Pointe\/} or Bernard Parmeggiani's {\em Ondes}. Nonetheless, transcribing it into standard notation is difficult, and instead, transcriptions may involve free shapes representing dynamics or timbre \citep{favreau2010acousmographe}.

From a commercial perspective, \textit{musique concr\`ete} is niche. The music charts indicate that the music that is most listened to is recent popular music. Popular music is distinguished from folk and art music by its use of recorded sound as the main mode of transmission \citep{tagg1982analysing,middleton1990studying,mazzanti2019defining}. It is characterised by commercial interests, entertainment, and ties to mass media \citep{frith2004popular}.

\citet{deruty2022development} define \textit{Contemporary Popular Music} (CPM) as current popular music, known for its technological innovation and cross-genre influences \citep{mazzanti2019defining, frith2004popular}. CPM genres include post-rock, rap/hip-hop, electronica, and non-Western styles like K-pop and reggaeton. \citet{bertin2011million} use the term, \textit{contemporary popular music}, to refer to tracks from 1922 onwards. We use it to refer to popular music since the 1970s, with \textit{Sgt. Pepper's Lonely Hearts Club Band} \citep{mccartney1997paul} as a milestone marking the beginning of the period. Like \textit{musique concr\`ete}, recent popular music is studio-produced. In \citepos{parry1911style} terms, both genres share the same `resources of the performance' (the studio), with similar constraints and affordances.

Has popular music abandoned musical notes? \citet{AIsongcontest_long} identify building blocks in popular music. The `melody' and `harmony' blocks suggest that popular music can be transcribed to musical notes, either automatically \citep{bertin2011million} or manually \citep{tagg1982analysing}. On the other hand, percussion, important enough for research in automatic transcription \citep{wu2018review} and musicology \citep{mowitt2002percussion}, is characterised by clear stochastic noise components in instruments like snare drums \citep[pp. 602-606]{fletcher2012physics}, making their transcription to musical notes difficult.

Percussion is not the only aspect of CPM that complicates transcription to musical notes. For example, \emph{sampling} introduces real-life noisy sounds into music \citep[p. 408]{forman2004s}, and the noise surrounding each partial in guitar distortion \citep[p. 184]{berger2005heaviness} can blur frequencies, making pitch identification difficult. The fact that not all popular music can be transcribed meaningfully into notes (i.e., staff notation) raises the question: can we quantify how much popular music deviates from being a combination of `notes'? In other words, to what extent does it deviate from being produced (or producible) by acoustic resonators or technology that emulates them? This article aims to begin answering this question. We hope this article provides the beginning of an answer to this question.



\section{Datasets, features, analysis}\label{sec:datasetsfeaturesanalysis}

\subsection{Datasets}\label{subsec:datasets}

In this study, we use the following four datasets of stereo audio tracks.\footnote{All tracks are sampled at 22050Hz.}

\begin{enumerate}
    \item The popular music dataset, or `BEA dataset', contains 30,435 tracks released between 1961 and 2022, with at least 460 tracks per year. It extends the dataset used by \citet{deruty2015mir}, and is based on the `Best Ever Albums' website \citep{besteveralbums}, which aggregates reviews to rank top-rated albums each year. As these ratings are dynamic, the dataset reflects a snapshot at the time of writing. Both original and remastered versions are included, and the influence of remastering is discussed in Section~\ref{sec:medium}.
    
    \newpage
    
    \item  The piano dataset contains 4600 piano audio tracks from various sources, including Alpha's `Schumann Project', Ciccolini's complete EMI recordings, and Brendel's complete Decca recordings. Tracks from this dataset range from the Viennese Classical era to the early 20\textsuperscript{th} century. Different interpretations of the same original score may be found in the dataset.
    \item The orchestra dataset contains 10800 orchestral and opera tracks from various sources, including Deutsche Grammophon's `Classical Gold', `The History of Classical Music', Decca's `55 Great Vocal Recitals' and `Ultimate Boxset' series. Tracks in this dataset range from the Baroque era to the early 20\textsuperscript{th} century. Different interpretations of the same original score may be found in the dataset.
    \item The \textit{musique concr\`ete} dataset consists of 1000 tracks from composers related to Pierre Schaeffer's school of thought by either having directly collaborated with him or having produced music at INA/GRM in Paris. It includes music by Pierre Henry, Bernard Parmegiani, Denis Dufour, and Fran\c{c}ois Bayle amongst others.
\end{enumerate}

\subsection{Features}\label{subsec:features}

One key aspect of the study is the quantitative evaluation of inharmonicity in music. Inharmonicity has been defined as `the departure in frequency from the harmonic modes of vibration' \citep{young1952inharmonicity} and `the deviation of a set of frequencies from an exact harmonic series' \citep{campbell2001inharmonicity}. A further distinction can be observed between `inharmonic sounds which have little if any relevance for music (e.g., white or pink noise)' \citep{schneider2009perception} and `coherent'  inharmonic signals, which `sound as stable and smooth as harmonic signals' \citep{deboer1956pitch}. In other words, inharmonicity can be the result of either inharmonic partial relations or noise (Appendix~\ref{subsubsec:HRandnoise} elaborates on the relationship between noise and inharmonicity). As a result, the features we use must address both partial relations and noise.

\subsubsection{HR-inharmonicity} \label{sec:HRinharmonicity}

We use the Matlab R2022b implementation of the MPEG-7 feature \textit{HarmonicRatio} as part of the broader inharmonicity analysis.\footnote{\url{https://fr.mathworks.com/help/audio/ref/harmonicratio.html}} \textit{HarmonicRatio} has been described as measuring `the proportion of harmonic components within the power spectrum' \citep[p. 36]{mpeg7audio} and `the ratio of harmonic power to total power' \citep[p. 33]{moreau2006mpeg}. To our knowledge, \textit{HarmonicRatio} is the only feature in the literature that measures the proportion of harmonic components in the general case. Features such as \textit{Inharmonicity} in \citet[p.17]{peeters2004large} measure `the [...] divergence from a purely harmonic signal' and can only be measured in relation to a single complex tone. 

\textit{HarmonicRatio} derives from the normalised auto-correlation of the signal. The normalisation is performed so that the auto-correlation at zero lag equals one. The feature output is the maximum value of the normalised auto-correlation after the first zero-crossing. \textit{HarmonicRatio} assumes values between 0 and 1, with harmonic complex tones resulting in 1. As we want to measure inharmonicity, we define a feature called {\em HR-inharmonicity}, defined as $1- HarmonicRatio$.  Appendix~\ref{sec:HRinterpretation} elaborates on the relationship between different aspects of the signal and HR-inharmonicity.

\newpage

\citet{patterson1996relative} and \citet{yost1996pitch,yost1997pitch} have argued that the `first peak of the auto-correlation function' can be used to evaluate the `saliency or the strength of pitch of complex sounds'---in other words, the perceptual impression of \emph{pitch strength}. The matter was further investigated by \citet{shofner2002pitch}. The illustrations in \citet[p. 3330]{yost1996pitch} and \citet[p. 439]{shofner2002pitch} show that the `first peak of the auto-correlation function' is the maximum value of the normalised auto-correlation after the first zero-crossing---that is, the same as what the MPEG-7 standard designates as \textit{HarmonicRatio} \citep{mpeg7audio,moreau2006mpeg}. One of our main concerns in this study is quantifying the degree to which a signal deviates from being harmonic. However, as \textit{HarmonicRatio} can be used to gauge the strength of pitch, the properties of the signal that we find to result in lower (or higher) \textit{HarmonicRatio} values may also lead to a corresponding decrease (or increase) in the perceived strength of pitch.

\subsubsection{Peak prominence and noisiness}\label{sec:noisinessintro}

Features have been proposed to measure the `noisiness' of the signal, such as spectral flatness \citep{peeters2004large} and \textit{AudioSpectrumFlatnessType} \citep{mpeg7audio}. However, these two features are sensitive to the overall envelope of the spectrum: music and pink noise result in comparable spectral flatness values, which defeats the purpose of measuring `noisiness'.  We introduce a new metric derived from spectral flatness and \textit{AudioSpectrumFlatnessType} that is robust to the overall spectral profile. We refer to this feature as {\em peak prominence\/} and to the inverse of peak prominence as {\em noisiness}. Appendix~\ref{sec:noisiness} details the elaboration of the metric.

\subsubsection{Inharmonicity}\label{sec:inharmintro}

HR-inharmonicity and peak prominence are not independent. Adding white noise to a harmonic complex tone does not change the partial positions, yet it lowers the maximum value of the normalised auto-correlation after the first zero-crossing, resulting in lower HR-inharmonicity values (see section~\ref{subsubsec:HRandnoise} for more details). To measure inharmonicity independently from `noisiness', we perform a PCA on the distribution of HR-inharmonicity and peak prominence values that generates a 2-dimensional representation in which we denote the dimensions by PC1 and PC2. PC1 represents the total amount of noise and inharmonicity, while PC2 represents the proportion of this total that is due to inharmonic intervals between partials. In this paper, we refer to PC2 as \textit{inharmonicity}.

Such an understanding of the term `inharmonicity' does not align perfectly with that of other authors. For some authors, `inharmonicity' designates the fact that one can observe overtones that deviate from an exact harmonic series \citep[e.g.,][]{young1952inharmonicity,campbell2001inharmonicity,klapuri2003multiple,micheyl2010pitch}. In this sense, the term does not reflect a judgment on the {\em amount\/} of deviation. \citet{fletcher1962quality}, among others, use an `inharmonicity coefficient' that does reflect such a judgment, but it remains specific to cases where the positions of overtones gradually diverge from harmonic relations as frequencies increase. \citet{schneider2009perception} extend the notion of `inharmonicity' to noisy signals but do not provide a feature to designate the prevalence of inharmonicity in the general case.

\subsubsection{Audio weighting}\label{sec:audioweightingintro}

\begin{figure}[h!]
  \centering
  \includegraphics[width=1\columnwidth]{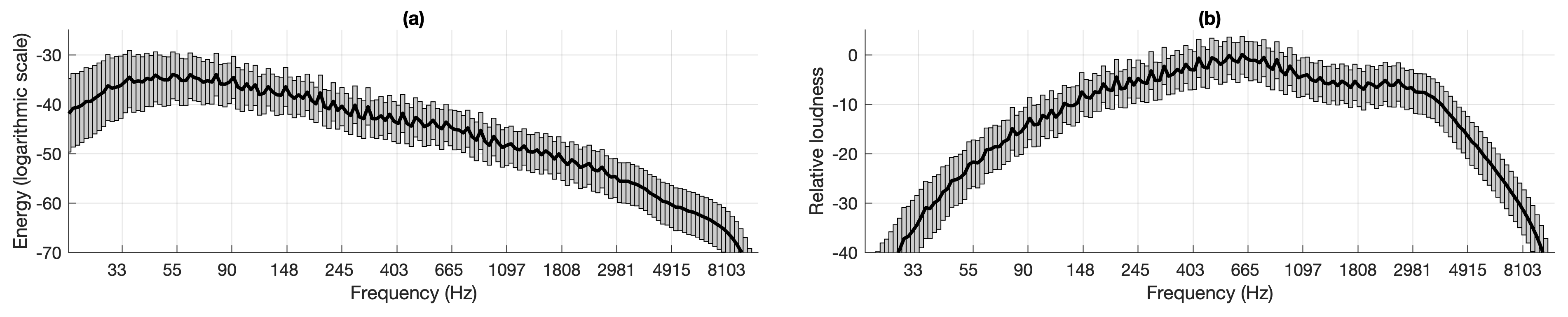}
  \caption{(a) power spectrum values for the BEA dataset. (b) relative loudness values.}
\label{fig:spectra_raw_weighted}
\end{figure}

Figure~\ref{fig:spectra_raw_weighted} (a) shows the power spectrum values for the BEA dataset. These results are consistent with those of \citet{pestana2013spectral}. However, these power spectrum values are not an accurate reflection of what the listener actually hears. Humans are not equally sensitive to all frequencies \citep{fletcher1933loudness}. Several models exist for the level of tones that are perceived as equally loud by human listeners depending on their frequency and sound pressure level. One such model is ISO226:2003 \citep{iso2262003}. Figure~\ref{fig:spectra_raw_weighted} (b) shows the power spectrum weighted using the ISO226:2003 equal-loudness contour corresponding to 50 phon, which is the median contour. Most notably, bass frequencies become attenuated. To more accurately reflect what the listener hears, we weigh the audio using the ISO226:2003 50 phon equal-loudness contour and evaluate the above features on both raw and weighted audio (see Appendix~\ref{sec:weighting} for more details).

\subsection{Corpus study}\label{sec:results}

\subsubsection{Measures of HR-inharmonicity}\label{subsec:HRmeasures}

\begin{figure}[h!]
  \centering
  \includegraphics[width=1\columnwidth]{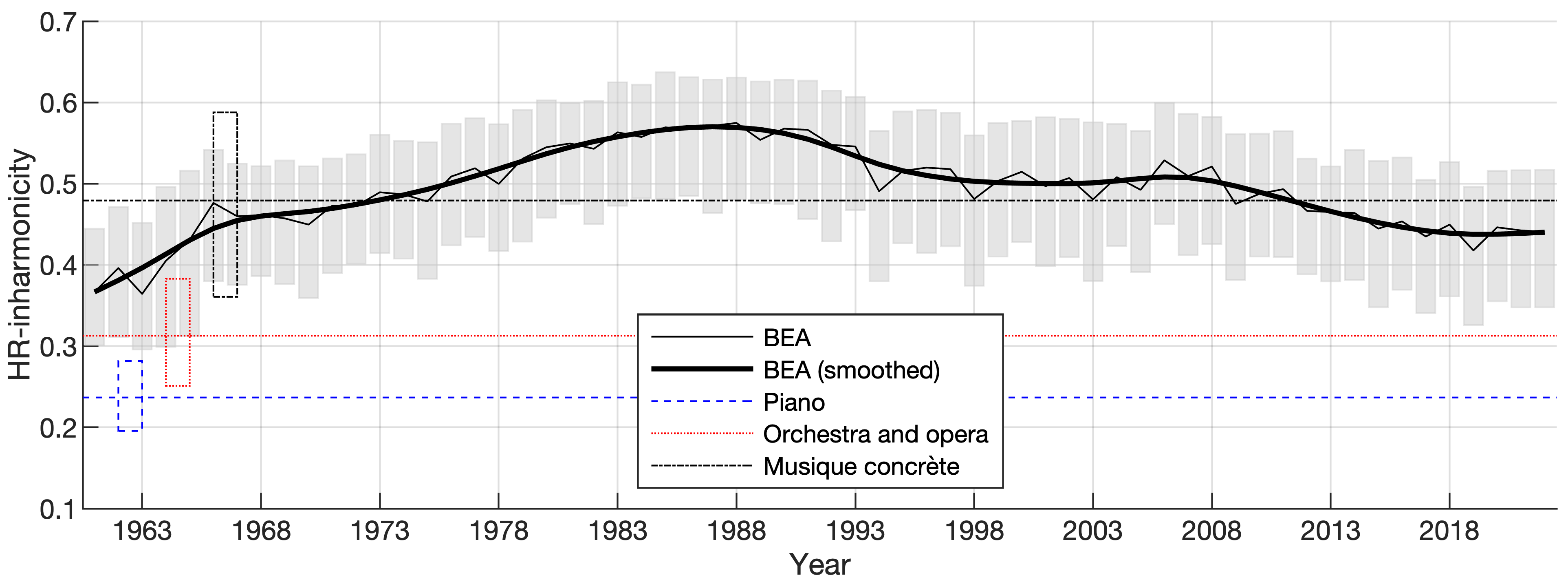}
  \caption{HR-inharmonicity for the four datasets. As in all the following similar figures, the lines represent median values, and the boxes represent interquartile ranges.}
\label{fig:harmonicratio1D}
\end{figure}

Figure~\ref{fig:harmonicratio1D} shows HR-inharmonicity for the four datasets. As autocorrelation is robust to level, the tracks are first gated to remove low-level parts (e.g., extensively in bonus tracks), which would result in very low feature values as a result of the analysis of the background noise. The gating process only retains the parts of the original audio whose RMS power is above -20dB after peak normalisation. From Figure~\ref{fig:harmonicratio1D}, it is possible to make the following observations.

\begin{enumerate}
    
    \item HR-inharmonicity of popular music is comparable to that of \textit{musique concr\`ete}. Both are more HR-inharmonic than piano and orchestral music. It is noteworthy that the two music categories that heavily use production techniques in the recording studio have similar HR-inharmonicity values.

    \item HR-inharmonicity of popular music is greater than that of orchestral music. Yet, orchestral music features a large number of instrumental sources, from typically 14--16 players during the second part of the 18\textsuperscript{th} century to 60--500 instrumentalists and voices during the 19\textsuperscript{th} century \citep{spitzer2001orchestra}. It suggests that greater HR-inharmonicity (lower pitch strength) in popular music may derive from other causes than the number of sources, such as the presence of less `pure' intervals and/or noise.
    
    \item Popular music evolution over the years shows an increase in HR-inharmonicity (a decrease in pitch strength) until the mid-1980s, followed by an irregular, slower decrease that has continued up to the present day.
    
\end{enumerate}

In Appendix~\ref{subsec:beating}, we point out that HR-inharmonicity is correlated with the music's `roughness' (in the sense of \citet{rasch1982perception} and \citet{masina2022dyad}).


\vspace{1cm}
\begin{figure}[htbp]
  \centering
  \includegraphics[width=1\columnwidth]{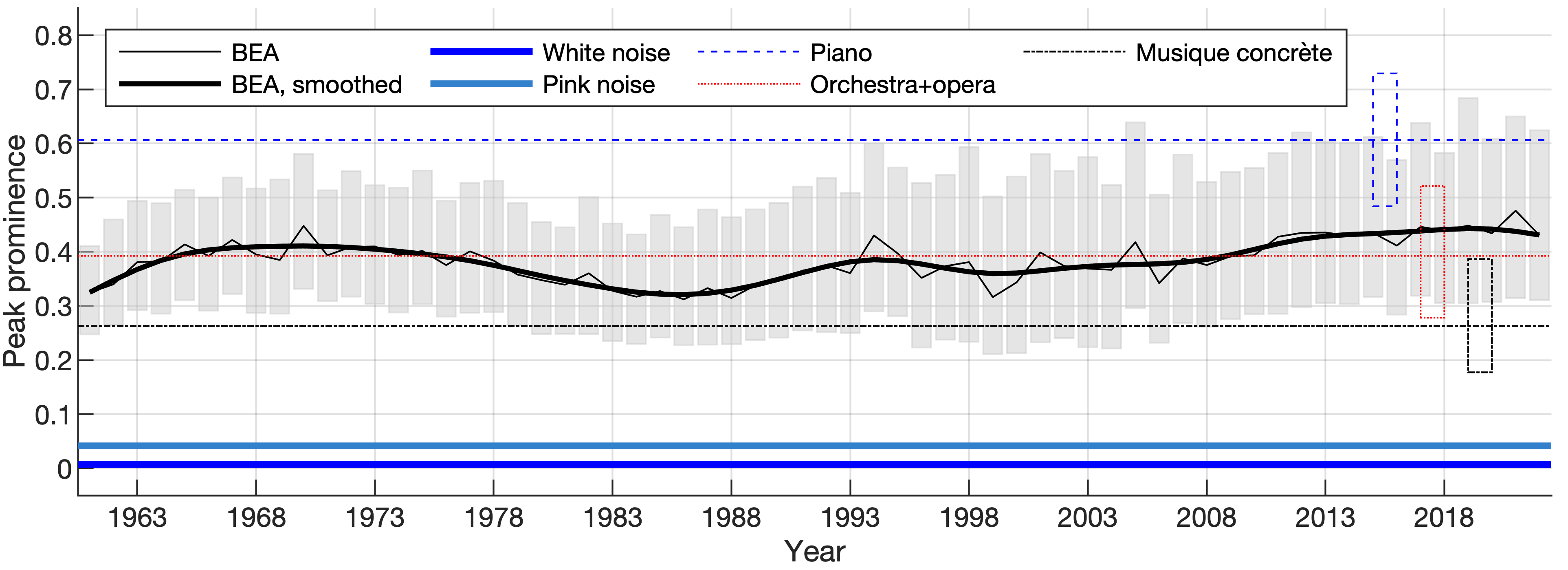}
  \caption{Measures of peak prominence for the four datasets, compared with values for pink noise and white noise.}
\label{fig:peakprominence1D}
\end{figure}

\subsubsection{Measures of peak prominence}\label{peakprominencemeasures}

Figure~\ref{fig:peakprominence1D} shows the peak prominence for the four datasets. From this graph, we can make the following observations.

\begin{enumerate}
    \item The values corresponding to white noise and pink noise are similar. The peak prominence values for white noise and pink noise are close to each other and clearly separated from music. It shows that peak prominence is robust to the global spectral envelope of the signal (which is not the case for spectral flatness as mentioned above).
    
    \newpage
    
    \item Popular music and orchestral music are similarly noisy. From Figure~\ref{fig:harmonicratio1D}, we derived that greater inharmonicity in popular music may stem from the use of inharmonic sources, more inharmonic musical intervals, and/or noise. We can now exclude noise from the causes. 
    The greater HR-inharmonicity observed in popular music is, therefore, likely to derive from a greater use of complex tones with inharmonic partials as well as the more frequent use of inharmonic intervals {\em between\/} complex tones.
    \item Piano music is less noisy than popular music, and \textit{musique concr\`ete} is noisier.
    \item In the case of popular music, there is a local maximum at around 1970, a local minimum during the mid '80s, and peak prominence increases slowly after 1986, continuing up to the present day. Notice that the peak prominence's interquartile range generally increases over the entire period considered.
\end{enumerate}

\subsubsection{HR-inharmonicity and peak prominence: BEA dataset}\label{subsec:BEA2D}

\begin{figure}
  \centering
  \includegraphics[width=1\columnwidth]{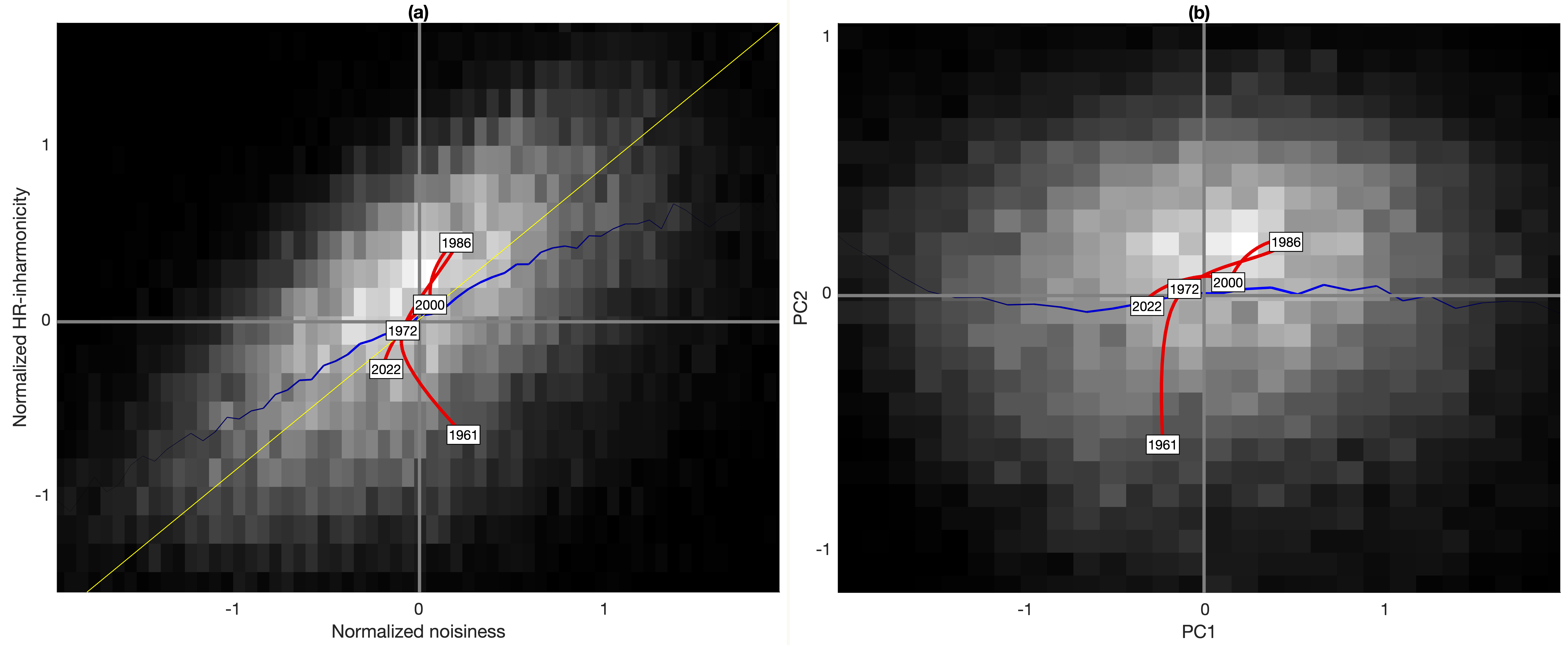}
  \caption{Noisiness and HR-inharmonicity in the BEA dataset. In (a), the $x$-axis is noisiness, while the $y$-axis is HR-inharmonicity. Distributions are made closer to normal by skewing ($x_{norm} = x^{0.18}$ and $y_{norm} = y^{0.21}$) and normalising (subtraction of the median and division by the interquartile range). The gray lines indicate the median values for each axis. The blue line indicates the median inharmonicity value for a given noisiness value. The yellow line shows the axis that explains the most variance in the distribution. The red curve shows how the centroid for the music in the dataset from a particular year evolves over the period 1961--2022, smoothed for readability purposes. The amount of smoothing is similar to that of the `smoothed' BEA curves in Figures \ref{fig:harmonicratio1D} and \ref{fig:peakprominence1D}. (b) shows the results of carrying out a PCA on the results shown in the graph on the left, with $x$-axis being PC1 and $y$-axis being PC2. The blue line indicates the median PC2 value for a given PC1 value, and the red curve shows how the centroid for a particular year evolves over the period 1961--2022.}
\label{fig:2Dchronologies}
\end{figure}

Figure~\ref{fig:2Dchronologies}, left, shows the result of the HR-inharmonicity vs. noisiness measures on the unweighted BEA dataset, including the evolution of both features according to the music's year of release. For a single HR-inharmonicity value (same vertical position), it is possible to distinguish between inharmonicity that arises from noise (right-hand side of graph) or intervals between discrete components (left-hand side of graph). In Figure~\ref{fig:2Dchronologies}, left, the bottom left corner of the graph corresponds to low noise and low HR-inharmonicity; whilst the top right corner corresponds to high noise and high HR-inharmonicity. The diagonal line extending from the bottom left of the graph to the top right corresponds to an increasing sum of noise and HR-inharmonicity. It is the axis that explains the most variance in the distribution.

Figure~\ref{fig:2Dchronologies}, right, shows the result of the distribution's PCA. In the upper two quadrants of the PCA graph, a relatively high proportion of the total amount of noise and inharmonicity is due to inharmonic intervals; while in the lower two quadrants, we have a relatively high proportion of the total amount of noise and inharmonicity being due to noise. Music tracks in the upper-right part of the PC2-vs.-PC1 representation feature the highest amount of inharmonic relations between partials (high total amount of noise and inharmonicity, low proportion of this being due to noise). Conversely, tracks in the lower-left part feature the lowest density of inharmonic relations between partials.

The graphs in Figure~\ref{fig:2Dchronologies} suggest that the evolution of popular music between 1961 and 2022 can roughly be divided into three eras as follows.

\begin{enumerate}

    \item {\bfseries 1961--1972$\quad$} During this period, the evolution is parallel to the PC2 axis. The total amount of noise and HR-inharmonicity remains roughly constant but the relative amount of HR-inharmonicity increases, indicating that higher HR-inharmonicity values are due to an increase in the use of inharmonic intervals between partials rather than an increase in noise. It also indicates a decrease in the relative amount of noise. The end of the time frame corresponds to the advent of extensive multi-tracking \citep{milner2011perfecting}. We take it as a starting date for music referred to as `Contemporary Popular Music' (CPM) in the sense of \citet{deruty2022development}, with the use of extensive multi-tracking and, more generally, heavy use of the recording studio.
    
     \newpage

    \item {\bfseries 1972--1986$\quad$} During this period, there is generally increasing noise and increasing HR-inharmonicity with a relatively slightly greater increase in HR-inharmonicity, indicating that a slightly greater proportion of the increasing HR-inharmonicity is due to inharmonic intervals, not noise. 
    
    \item {\bfseries 1986--2022$\quad$} During this period, the trend of the previous period is reversed and extends even beyond the values at the end of the first period (i.e., the value for 1972). During the period 1986--2022 there was a general decrease in the total sum of noise and HR-inharmonicity, with music of today having, on average, roughly the same total amount of noise and HR-inharmonicity as music from the early 1960s. However, the amount of HR-inharmonicity has fallen more than the amount of noise, meaning that the proportion of HR-inharmonicity in the sum of HR-inharmonicity and noise (i.e., the PC2 value) has slightly fallen. In other words, the music of today has roughly the same total amount of noise and HR-inharmonicity as the music from the early 1960s, but a much higher proportion of this sum is due to HR-inharmonicity caused by inharmonic intervals rather than noise.
\end{enumerate}

\begin{figure}[htbp]
  \centering
  \includegraphics[width=1\columnwidth]{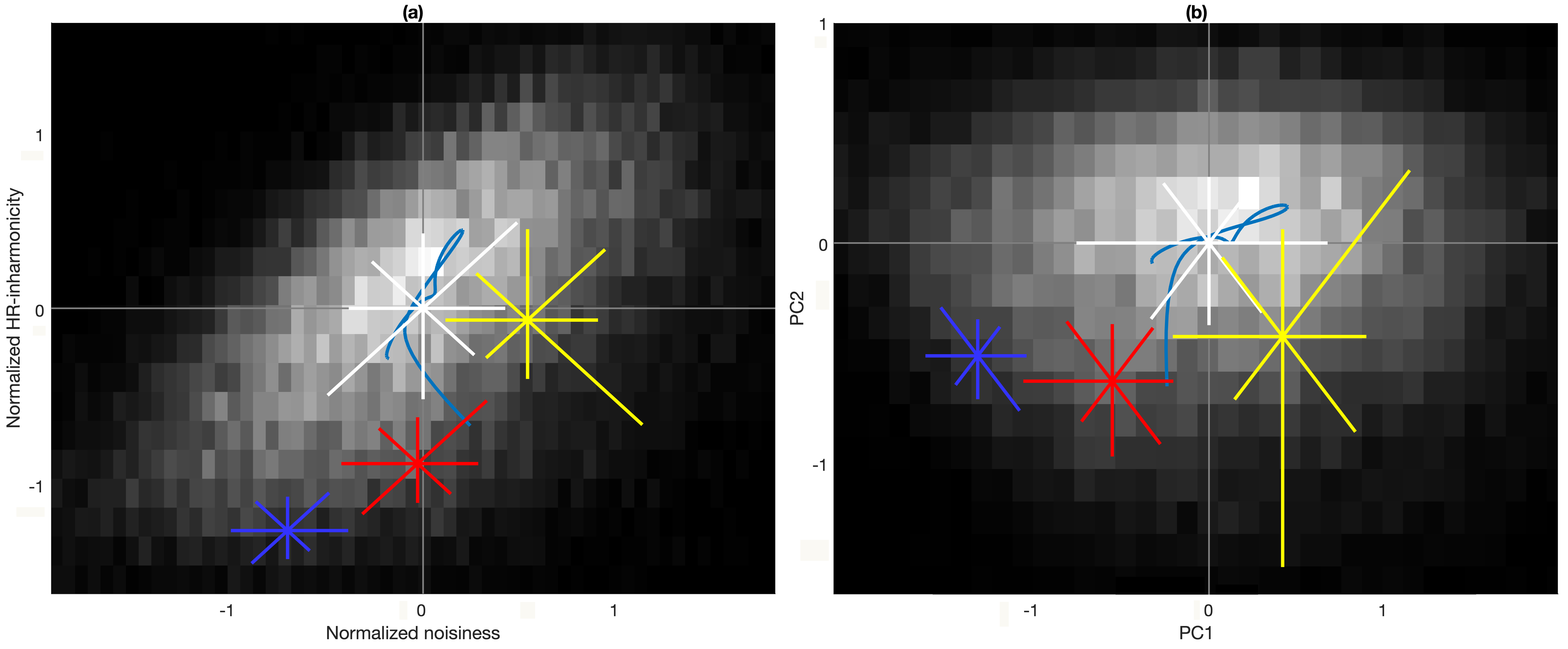}
  \caption{Respective positions of the four datasets. (a) HR-inharmonicity and noisiness. (b) PC1 and PC2 (as in Figure~\ref{fig:2Dchronologies}). The white, red, blue, and yellow stars respectively correspond to the BEA, orchestra, piano, and \textit{musique concr\`ete} datasets. The stars' lines start at the median coordinates for each dataset, and stop where half of the points starting from the median were met. The aquamarine-coloured curves at the centres of the images replicate the red curves in Figure~\ref{fig:2Dchronologies}.}
\label{fig:2Dfourdatasets}
\end{figure}

\subsubsection{HR-inharmonicity and peak prominence: all datasets}\label{sec:alldatasets}

Figure~\ref{fig:2Dfourdatasets} compiles the distribution values for the four datasets. As the distributions involving the four datasets are multimodal, we do not re-evaluate the PCs on the entire data, but keep the BEA's PCs as axes. From Figure~\ref{fig:2Dfourdatasets}, we can observe the following:

\begin{enumerate}

    \item Popular music from the beginning of the 1960s shows feature values that are not too different from that of orchestral music. However, CPM (popular music after 1972) uses more inharmonic intervals.

    \item The ratio of HR-inharmonicity to noise is similar for piano music, orchestral music and \textit{musique concr\`ete}. However, the total amount of HR-inharmonicity and noise increases from piano music to orchestral music to \textit{musique concr\`ete}. The ratio of HR-inharmonicity to noise is higher for CPM than the other genres, indicating that relatively more of the HR-inharmonicity in CPM is due to inharmonic intervals.

    \item As we discussed in Section~\ref{subsec:extent}, \textit{Musique concr\`ete} and CPM use similar `resources of the performance' \citep{parry1911style}, that is to say, they are both typically created in music production studios using electronic technology. In \citepos{mcpherson2020idiomatic} terms, \textit{musique concr\`ete} explores a noisier `space for musical exploration' with fewer inharmonic intervals between partials, and CPM explores a less noisy space with more inharmonic intervals. Both are equally distant from orchestral music.

\end{enumerate}

\subsubsection{HR-inharmonicity and peak prominence, weighted audio}\label{sec:HRpeakweighted}

\begin{figure}[htbp]
  \centering
  \includegraphics[width=1\columnwidth]{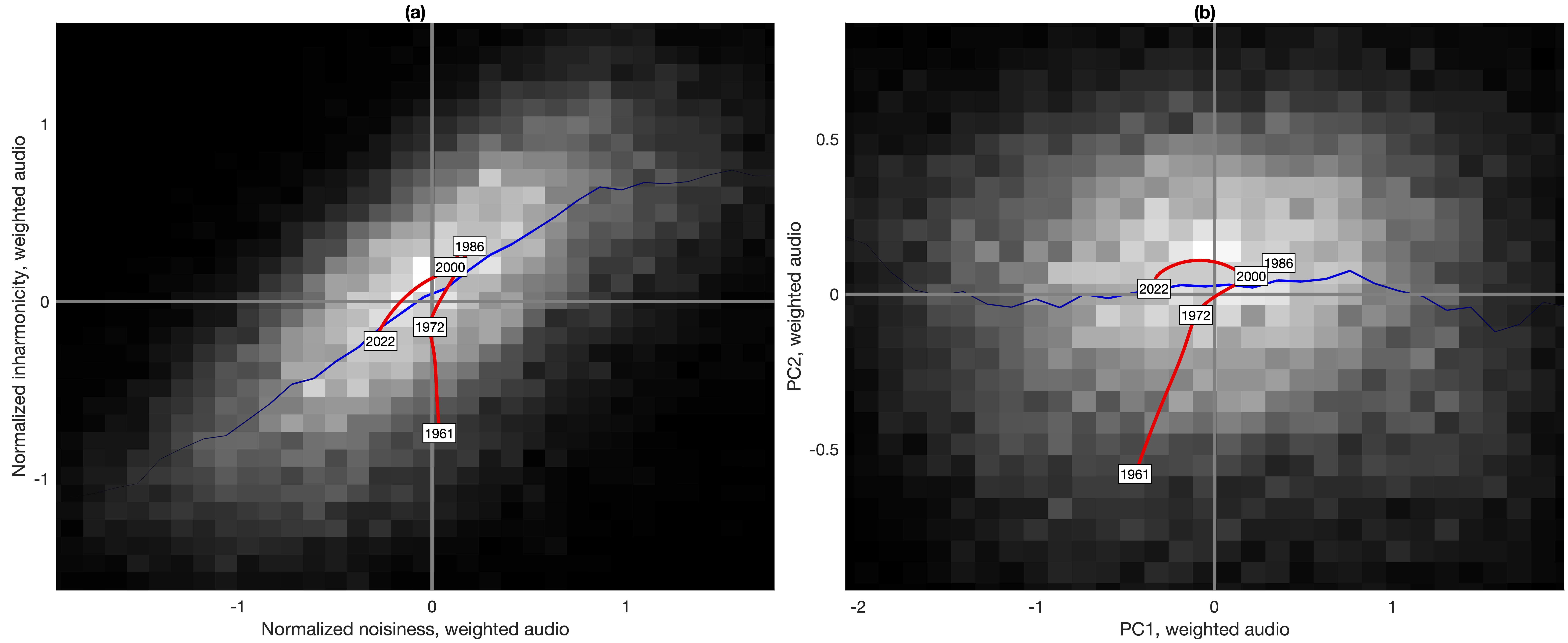}
  \caption{BEA dataset. (a) Normalised noisiness and inharmonicity from weighted audio. As in Figure~\ref{fig:2Dchronologies}, distributions are made closer to normal by skewing ($x_{norm} = x^{0.39}$ and $y_{norm} = y^{0.14}$) and normalising (subtraction of the median and division by the interquartile range). (b) PC1 and PC2 from weighted audio.}
\label{fig:2Dchronologiesweighted}
\end{figure}

Figure~\ref{fig:2Dchronologiesweighted}~(a) shows the evolution of HR-inharmonicity and noisiness from weighted audio. Figure~\ref{fig:2Dchronologiesweighted}~(b) shows the evolution of PC1 and PC2 from weighted audio. Figure~\ref{fig:2Dchronologiesweighted} may better reflect what is actually heard than Figure~\ref{fig:2Dchronologies} (unweighted audio) while minimizing objective aspects of the signal that may be less perceptually salient, such as high energy values at the bottom end of the spectrum. The observable differences between Figures~\ref{fig:2Dchronologies} and \ref{fig:2Dchronologiesweighted} may be summarised as follows.

\begin{enumerate}
    \item The increase in the ratio of interval inharmonicity to noisiness (PC2) from 1961 to 1972 in Figure~\ref{fig:2Dchronologies} is accompanied by a noticeable increase in total inharmonicity and noise (PC1) in Figure~\ref{fig:2Dchronologiesweighted}~(b).
    
    \clearpage
    
    \item Between 1986 and 2000, the fall in inharmonicity to noise ratio (PC2) seems less than it does when viewed in terms of unweighted audio. It suggests that the faster evolution witnessed from non-weighted audio derives from very low frequencies. In other words, the faster evolution originates from a change in the properties of the {\em medium\/} in addition to that of the {\em audible content}. The weighting function gives relatively much less weight to the lower frequencies, meaning that the decrease in lower frequency energy between 1986 and 2000 would have less of an effect on the weighted audio graph of PC2 against PC1 (Figure~\ref{fig:2Dchronologiesweighted}~(b)). 
 
\end{enumerate}

\vspace{1cm}

\begin{figure}[htbp]
  \centering
  \includegraphics[width=1\columnwidth]{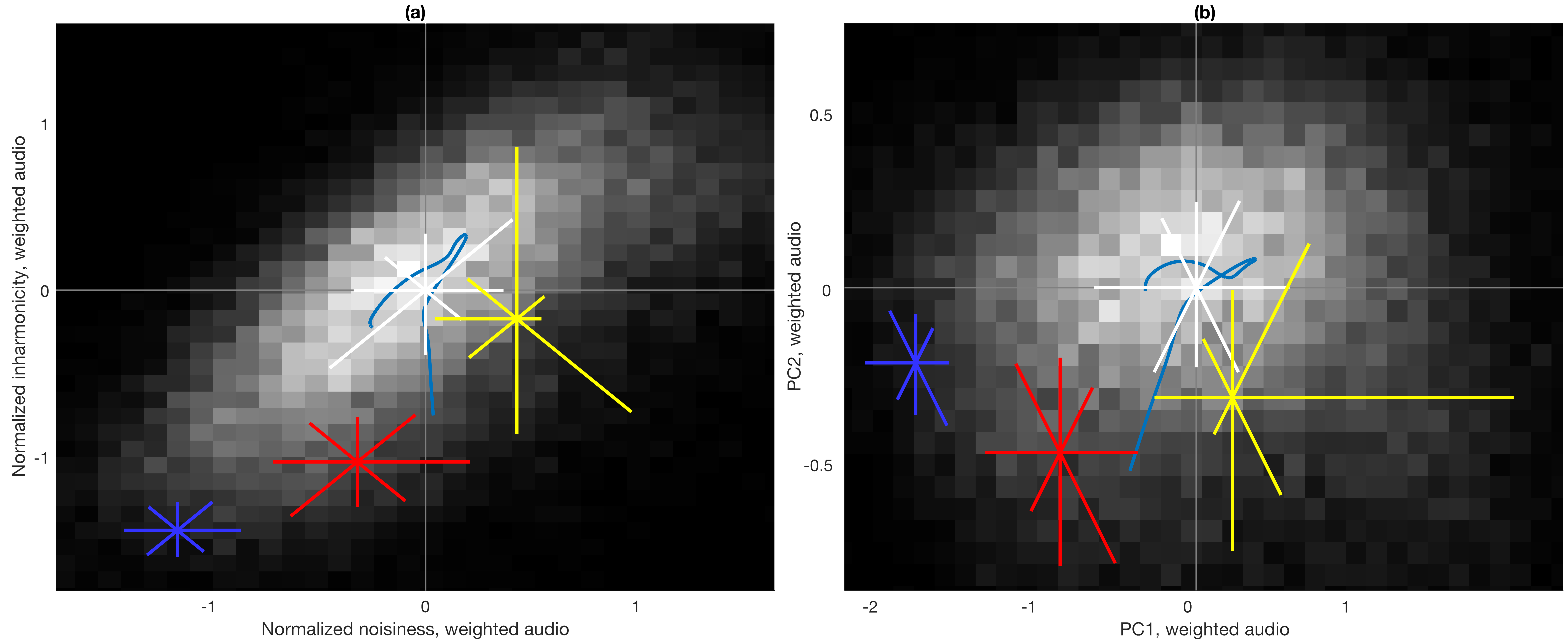}
  \caption{Respective positions of the four datasets using weighted audio. (a) Inharmonicity vs.~noisiness; (b) PC2 vs.~PC1. The white, red, blue and yellow stars correspond, respectively, to the BEA, orchestra, piano and \textit{musique concr\`ete} datasets, as in Figure~\ref{fig:2Dfourdatasets}.}
\label{fig:2Dfourdatasetsweighted}
\end{figure}



\newpage

\subsection{Relations with some other diachronic studies}\label{sec:diachronic}

\citet{mauch2015evolution} investigate the US Billboard Hot 100 between 1960 and 2010. Using music information retrieval and text-mining tools, they demonstrate quantitative trends in their harmonic and timbral properties. Using NNLS Chromas \citep{mauch2010approximate} and MFCCs, they find that `1964, 1983 and 1991 are periods of particularly rapid musical change'. They remark that `other measures may give different results', which is indeed the case in this study, where we observe the fastest changes occurring throughout the 1960s (see Figures~\ref{fig:2Dchronologies} and \ref{fig:2Dchronologiesweighted}), as well as turning points in 1972, 1986 and 2000. The features used in \citet{mauch2015evolution} do not seem well-correlated to the features we use, suggesting that chromas and MFCC may fail to recognise certain properties of the music. As a general rule, it may be that the interpretation of feature-based studies should be limited to the interpretation of the features themselves rather than to `the music' as a whole.

\vspace{1cm}

\begin{figure}[htbp]
  \centering
  \includegraphics[width=1\columnwidth]{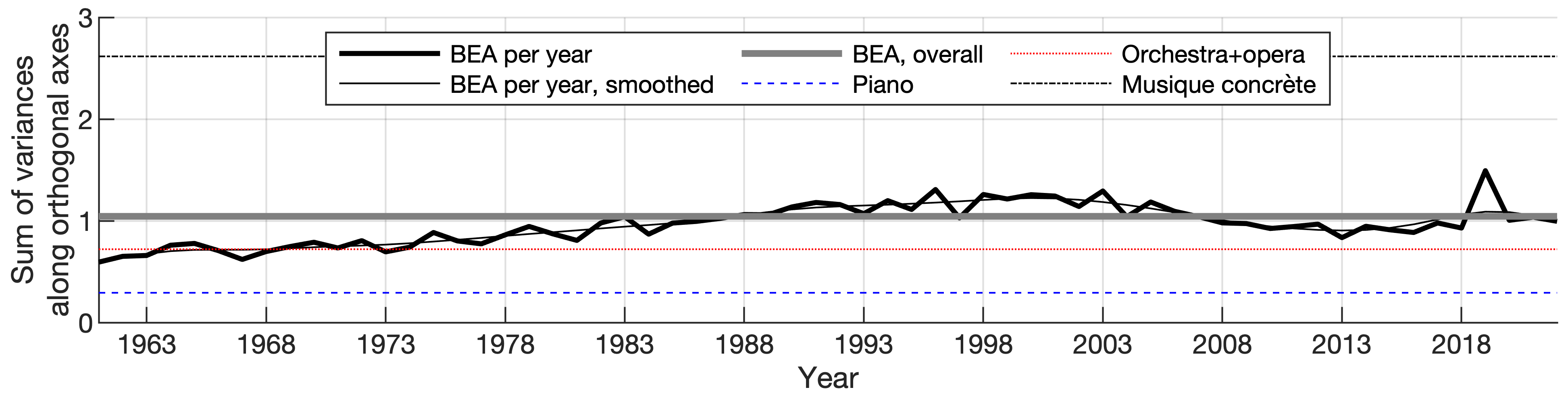}
  \caption{Variance of the distribution along PC1 and PC2 (unweighted audio).}
\label{fig:variances}
\end{figure}

In Figure~\ref{fig:2Dchronologies}, PC1 and PC2 are orthogonal. Therefore, the variance for the 2D data can be estimated as the sum of the variances along each axis. Figure~\ref{fig:variances} represents the sum of the variances along PC1 and PC2 for each dataset, as well as for the year-by-year data of the BEA dataset. The results are identical in terms of comparison for the original and PC representations. The following points are worth noting.

\begin{enumerate}

   \item The variance of popular music in terms of the features we use is higher than that of orchestral music after 1974, and higher overall. It suggests that in terms of noise and HR-inharmonicity, popular music uses a wider `space for musical exploration' than orchestral music.

   \item The variance of \textit{musique concr\`ete} in terms of the features we use is greater than that of popular music. In terms of noise and HR-inharmonicity, \textit{Musique concr\`ete} takes greater advantage of the lack of constraints in the production process.

    \item  \citet{serra2012measuring} find that the evolution of popular music goes `towards a consistent homogenization of the timbral palette'. Figure~\ref{fig:variances} suggests that no such homogenization of the timbral palette has taken place in terms of noisiness and inharmonicity (or, equivalently, peak prominence and pitch strength).
    
\end{enumerate}


\newpage
\subsection{Influence of the medium and remastering processes}\label{sec:medium}

The medium on which the music is recorded may contribute to noise in the recording. According to \citet{brandt2019automatic}, early recording media, such as wax cylinders, had signal-to-noise ratios (SNRs)  of below 40 dB. Vinyl discs have SNRs of 55 to 60 dB, and magnetic tape storage SNRs of 60 to 70dB. To evaluate the extent to which the medium contributes to noisiness, we compute the difference between the noisiness in the original BEA files and the same files on which we add white noise corresponding to an SNR of 40dB. The median of the absolute value of the difference is 0.000007. In Figure~\ref{fig:peakprominence1D}, typical noisiness values for the BEA dataset are shown to be approximately $0.3$. The comparison suggests that the contribution of the medium to the total amount of noise is negligible.

Music albums may be remastered. \citet{deruty2011dynamic} suggests that remastering significantly affects the album's loudness. We compared HR-inharmonicity values for the songs from the original and remastered version of the British band, The Cure, as \citet{deruty2011dynamic} did for loudness values. The median of the absolute value of the difference is 0.015 for non-weighted audio and 0.008 for weighted audio. In Figure~\ref{fig:harmonicratio1D}, typical HR-inharmonicity values for the BEA dataset range between 0.3 and 0.63. Along with the influence of the background noise, the comparison suggests that the study's results are robust to remastering processes.


\section{Analysis of specific tracks and artists}\label{sec:confrontation}

In this section, we focus on the music of particular artists using the analysis of weighted audio proposed in Section~\ref{sec:HRpeakweighted}. We suggest links between the analysis results and production methods.

\subsection{Selected artists}\label{subsec:selectedartists}

Figure~\ref{fig:Weighted_Examples_Straight} illustrates noisiness and HR-inharmonicity values for several artists and speech tracks (weighted audio). Artists were chosen that lie on the edges of the distribution so that it is easier to understand the perceptual meaning of the two dimensions. Figure~\ref{fig:Weighted_Examples_Rotated} shows the same data after PCA. We go through each element in Figures~\ref{fig:Weighted_Examples_Straight} and \ref{fig:Weighted_Examples_Rotated} so as to identify links between the feature values and aspects of the corresponding audio content.

\begin{enumerate}

    \item \textbf{Initial reference: speech tracks.} The speech tracks in Figures \ref{fig:Weighted_Examples_Straight} and \ref{fig:Weighted_Examples_Rotated} (white dots) are `interludes' or `skits' as found in hip-hop music. They generally have a low PC2 value. Low PC2 values for speech tracks illustrate how a sound can be non-inharmonic while not featuring stable pitch values. Higher PC2 values are observed in some of the examples of distorted voices. Higher PC1 values are observed in the examples that feature a high background noise.
    
    \item \textbf{PC2 values that are comparable to speech.} The two artists whose music corresponds to PC2 values similar to those for speech are Barbara and Judy Garland. Their music features a monodic lead singer with piano and/or orchestral accompaniment. Drums may be present in Garland's music but not in Barbara's, which may account for the lower noisiness and PC1 values of the latter.

\begin{figure}[H]
  \centering
  \includegraphics[width=1\columnwidth]{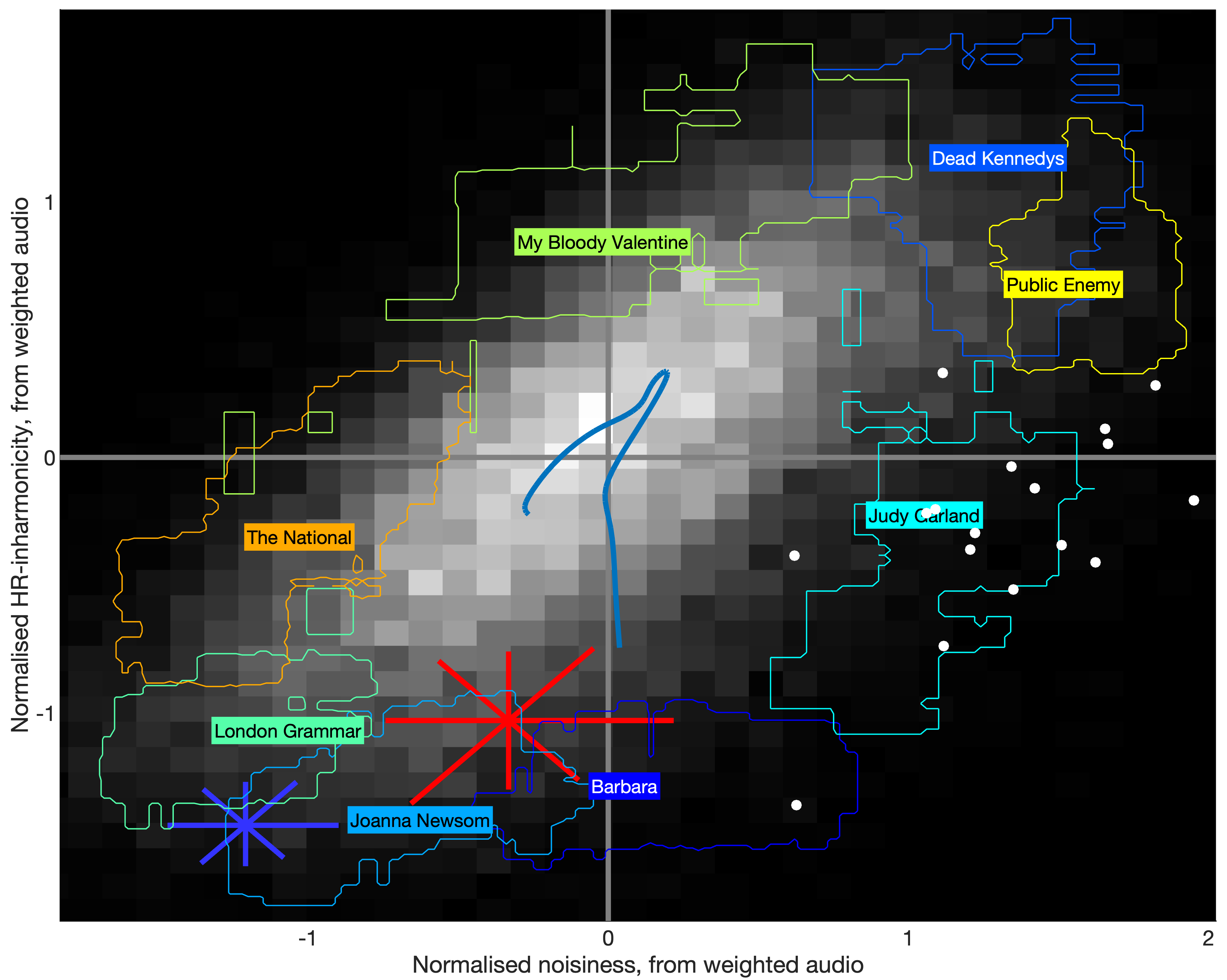}
  \caption{BEA dataset, normalised noisiness and HR-inharmonicity corresponding to particular artists as well as to monodic speech tracks. For each artist, the contours represent the isolines at half-height. Contours are evaluated on a mean filter of the distribution for better readability. The artist's name is centered on the median. The white dots correspond to individual monodic and possibly processed speech tracks (the `skits'). The red and blue stars denote the feature distributions for the orchestral and piano music datasets, as in Figure~\ref{fig:2Dfourdatasets}.}
\label{fig:Weighted_Examples_Straight}
\end{figure}

    \item \textbf{Exemplification of the 1961--1986 evolution.} The combined evolution of PC1 and PC2 from 1961 to 1986 is exemplified by the productions of Barbara and My Bloody Valentine. My Bloody Valentine's music is characterised by complex guitar textures with open tuning \citep{leonard2021bloody}, pitch bending, tremolo \citep{diperna1992bloody}, and an extensive effects rig \citep{double1992bloody}. Although chords can be identified, individual note perception is difficult, consistent with (a) the high number of inharmonic relations in their tracks (top right of the PC1--PC2 representation) and (b) the idea that pitch perception in inharmonic sounds is more ambiguous than in harmonic ones \citep{schneider2000inharmonic,schneider2009perception}. According to music critic Anthony Fantano, the guitar in their album `Loveless' is `slathered with waves of pink smog', obscuring and transforming the perception of the guitar \citep{fantano2013loveless}. The examples of Barbara and My Bloody Valentine suggest that, between 1961 and 1986, recorded music shifted from acoustic, tonal instruments playing content that could be transcribed to a score, to heavily processed, noisy, and inharmonic studio-made content, including drums.

\begin{figure}[H]
  \centering
  \includegraphics[width=1\columnwidth]{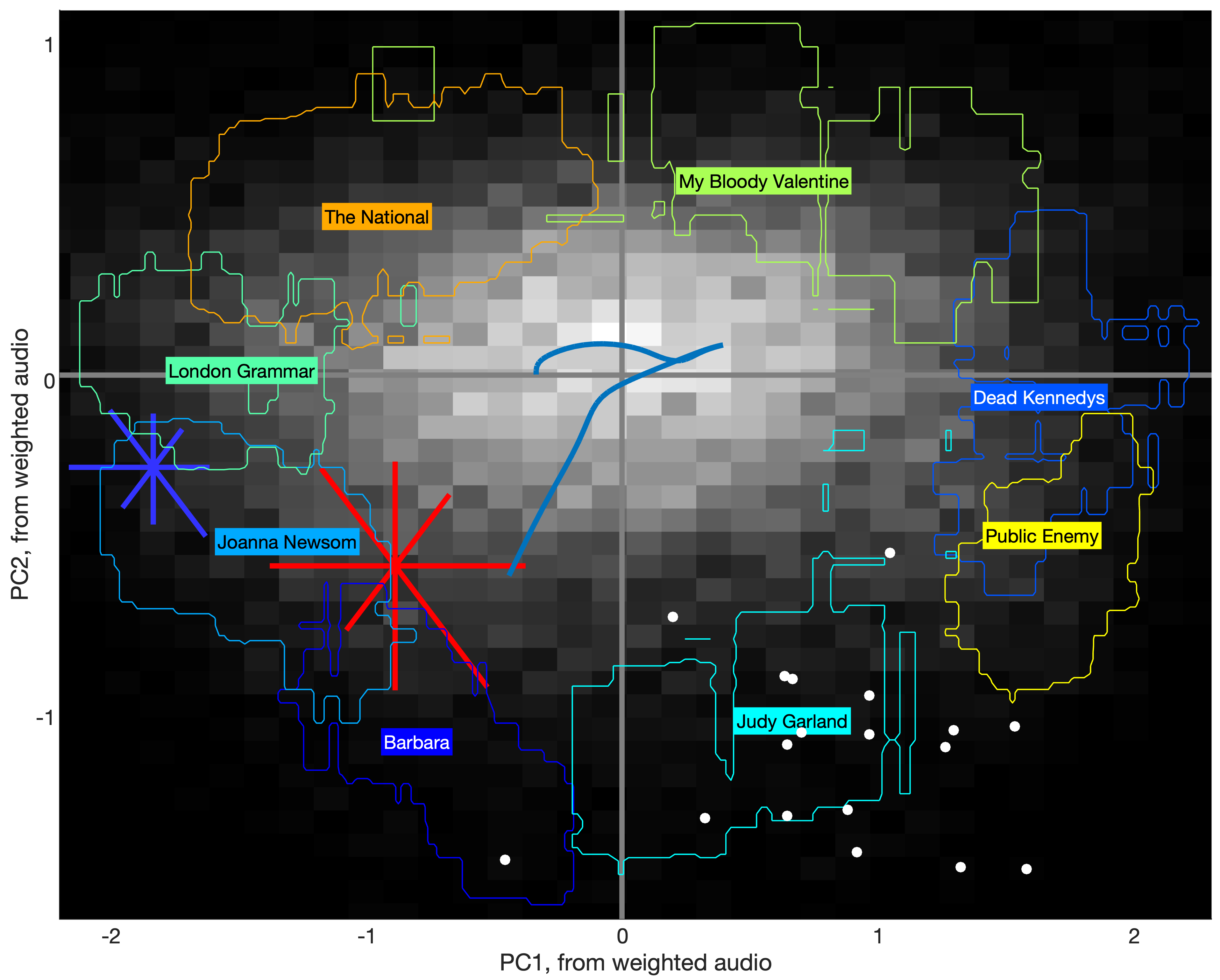}
  \caption{Same data as in Figure~\ref{fig:Weighted_Examples_Straight}, PCA.}
\label{fig:Weighted_Examples_Rotated}
\end{figure}

    \item \textbf{Noise from distortion.} The position of Dead Kennedys in Figures~\ref{fig:Weighted_Examples_Straight} and \ref{fig:Weighted_Examples_Rotated} shows that they exhibit very high total noise and inharmonicity, with the proportion of inharmonicity relative to noise being around the median for that level of noise. As a punk rock band from the 1980s, Dead Kennedys' music includes noisy vocals, drums, bass, and distorted guitar. The noise likely stems from the vocals, drums, and guitar. Drums cover a wide frequency range, and in heavy-metal-style distortion, noise surrounds each harmonic rather than layering with them \citep[p. 184]{berger2005heaviness}. Figure~\ref{fig:Dead_Kennedys} illustrates that the noise in their music is spread across different frequencies. Their PC2 values are lower than My Bloody Valentine's, likely due to the lower volume of guitar parts in the mix and fewer apparent layers.

        \item \textbf{Exemplification of the 1986--2020 evolution.} The combined evolution of PC1 and PC2 from 1986 to 2020 is illustrated by the productions of Dead Kennedys and London Grammar. Both bands have an inharmonicity-to-noise ratio near the BEA dataset median, but the total amount of noise and inharmonicity is very low for London Grammar and very high for Dead Kennedys. London Grammar, an indie pop band from the 2010s, creates studio- and synth-oriented music \citep{musictech2023londongrammar}, with carefully produced, atmospheric tracks \citep{soundonsound2014londongrammar,7digital2013londongrammar}. Their music features lead vocals with smoothed plosives, noiseless yet rich multi-layered arrangements, and heavy reverb. These layers and reverb likely explain the higher PC2 values compared to Joanna Newsom's music. The examples of Dead Kennedys and London Grammar suggest that, between 1986 and 2020, recorded music shifted from simpler, noisier tones to more polished, multi-layered studio productions. However, this evolution is less pronounced than the shift observed between 1961 and 1986.

\begin{figure}[H]
    \centering
    \includegraphics[width=1\columnwidth]{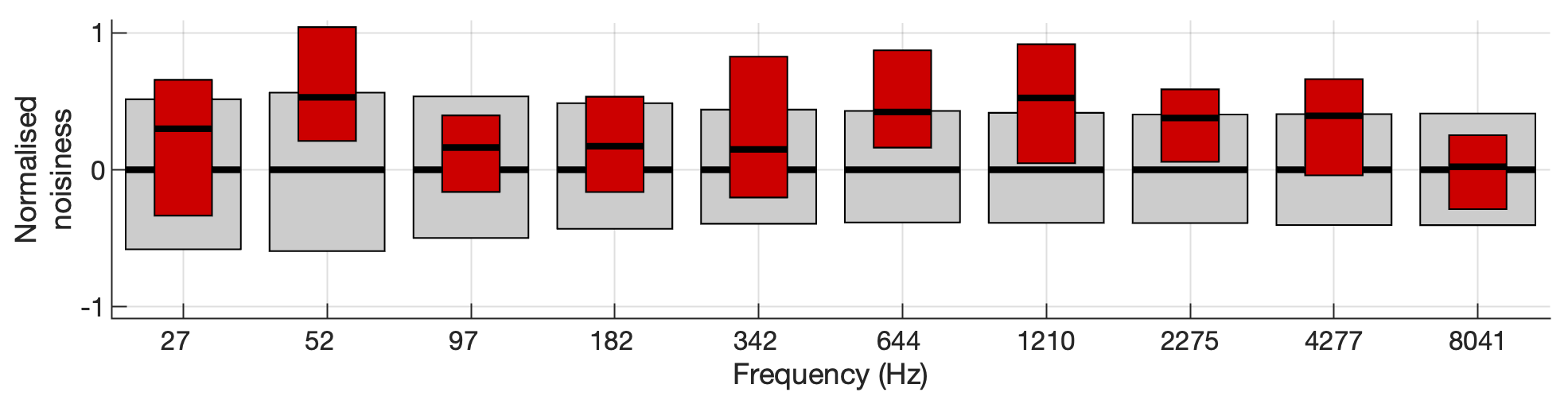}
    \caption{Band-by-band normalised noisiness for the Dead Kennedys' music. The gray rectangles represent the noisiness values for the entire dataset (25\textsuperscript{th}, 50\textsuperscript{th} and 75\textsuperscript{th} percentiles), adjusted so that the median is zero. The red rectangles represent the adjusted noisiness values for Dead Kennedys.}
    \label{fig:Dead_Kennedys}
\end{figure}

\vspace{1cm}
    
    \item \textbf{Slightly inharmonic music based on acoustic instruments.} Joanna Newsom's music has a higher PC2 value than Barbara's, but lower total noise and inharmonicity, indicating a higher inharmonicity-to-noise ratio in Newsom's music. This ratio stems from inharmonic tones. Newsom's music, featuring fewer layers than London Grammar's, suggests that fewer elements may result in less inharmonicity (see Appendix~\ref{subsubsec:severalcomplextones} for further discussion). Her music consists mainly of vocals accompanied by harp, with occasional piano and orchestra. The harp's difficulty in tuning \citep{cathcart2018harp} and its initial `twang' with higher frequencies converging to the pitch \citep{fletcher2000inharmonicity} may explain the higher PC2 values. The inharmonicity of piano strings \citep{rasch1985string} also aligns Newsom's music with piano dataset values. However, the PC2 value remains below the median, meaning the inharmonicity relative to noise is lower than average. Low noisiness likely results from clean instrumentation, minimal use of drums, and the absence of prominent plosives in the vocals.
    
    \item \textbf{Noise from sampling.} Public Enemy's music is as noisy as Dead Kennedys' but less inharmonic. The two Public Enemy albums in the dataset, released in 1988 and 1990, are characterised by the recombination of numerous samples across multiple layers \citep[pp. 22-26]{mcleod2011creative}. The sound has been described as `part musique concrète, [...] a noisy collage of sputtering Uzis, wailing sirens, fragments of radio and TV commentary [...], all riding on rhythms articulated by constantly changing drum voices [...], off-kilter loops, aliased or scratchy samples, and high-pitched spiraling sounds' \citep[p. 408]{forman2004s}, explaining the high noisiness. The lower inharmonicity-to-noise ratio compared to Dead Kennedys may stem from Public Enemy using fewer pitched elements. Like Joanna Newsom's music, Public Enemy's music has a lower-than-average inharmonicity-to-noise ratio (PC2). However, unlike Newsom's music, the total amount of inharmonicity and noise in Public Enemy's music is very high (PC1).

\newpage

    \item \textbf{Does studio work lead to more inharmonic partials?} The National's music is less noisy than My Bloody Valentine's, with a similar proportion of inharmonicity deriving from partials. Like My Bloody Valentine, The National's production involves extensive studio experimentation \citep{doyle2017national}. The band has two guitarists who use large pedal-boards, focusing on `textural soundscaping' rather than virtuosity. One of their goals is to create a `lattice work of notes' \citep{guitar2009national}. An example of this can be found in the album `Sleep Well Beast', where `the pianos are playing off each other by an eighth note' \citep{guitar2009national}, increasing the number of simultaneous pitch values. Since more distinct tones result in greater inharmonicity, such techniques likely contribute to higher inharmonicity values. Additionally, much sustain and reverb will also lead to increased inharmonicity because of tones that are close together in pitch in the same part or voice overlapping in time.

\end{enumerate}

A key observation from the above is that low PC2 values appear to correspond to tonal music with clearly distinguishable elements (Barbara, Judy Garland). If we consider the gradation from low to high PC2 values on the low PC1 side (Barbara, Joanna Newson, London Grammar, The National), then the music appears to gradually involve more and more studio work. This would result in keeping the total amount of inharmonicity and noise roughly constant (and low) and increasing the ratio of inharmonicity to noise, which means increasing the extent to which inharmonicity results from partials rather than noise. High PC2 values seem to correspond to music in which heavy studio production work is performed on `pitched' instruments, especially on guitars (The National, My Bloody Valentine). 

\subsection{Possible causes for high PC1 and PC2 values}\label{subsec:causes}

PC1 values are generally higher in the case of popular music than in the case of orchestral music (see Figures~\ref{fig:2Dfourdatasets} and \ref{fig:2Dfourdatasetsweighted}, as well as Figure~\ref{fig:PCAs1D} below). Possible factors for higher PC1 values may involve loud drums (Dead Kennedys, Public Enemy), distortion (Dead Kennedys, My Bloody Valentine), noisy samples (Public Enemy), and vocals with loud plosives (Dead Kennedys, Public Enemy).

PC2 values are also generally higher in the case of CPM than in the case of orchestral music. As PC2 is the ratio of inharmonicity to noise, a higher PC2 value indicates that noise makes a relatively smaller contribution to the total sum of inharmonicity and noisiness (i.e., to PC1). As previously stated, lower \textit{HarmonicRatio} values may be obtained either from properties deriving from each complex tone (e.g., inharmonicity) or from properties deriving from the combination of complex tones (e.g., number of sources and scales). Judging from the facts that (1) orchestras have a high number of sources, and (2) harmony in CPM does not appear to be more chromatic than that of Western classical music, we might conclude that the higher PC2 values in CPM originate from tone inharmonicity.

\begin{figure}[htbp]
  \centering
  \includegraphics[width=1\columnwidth]{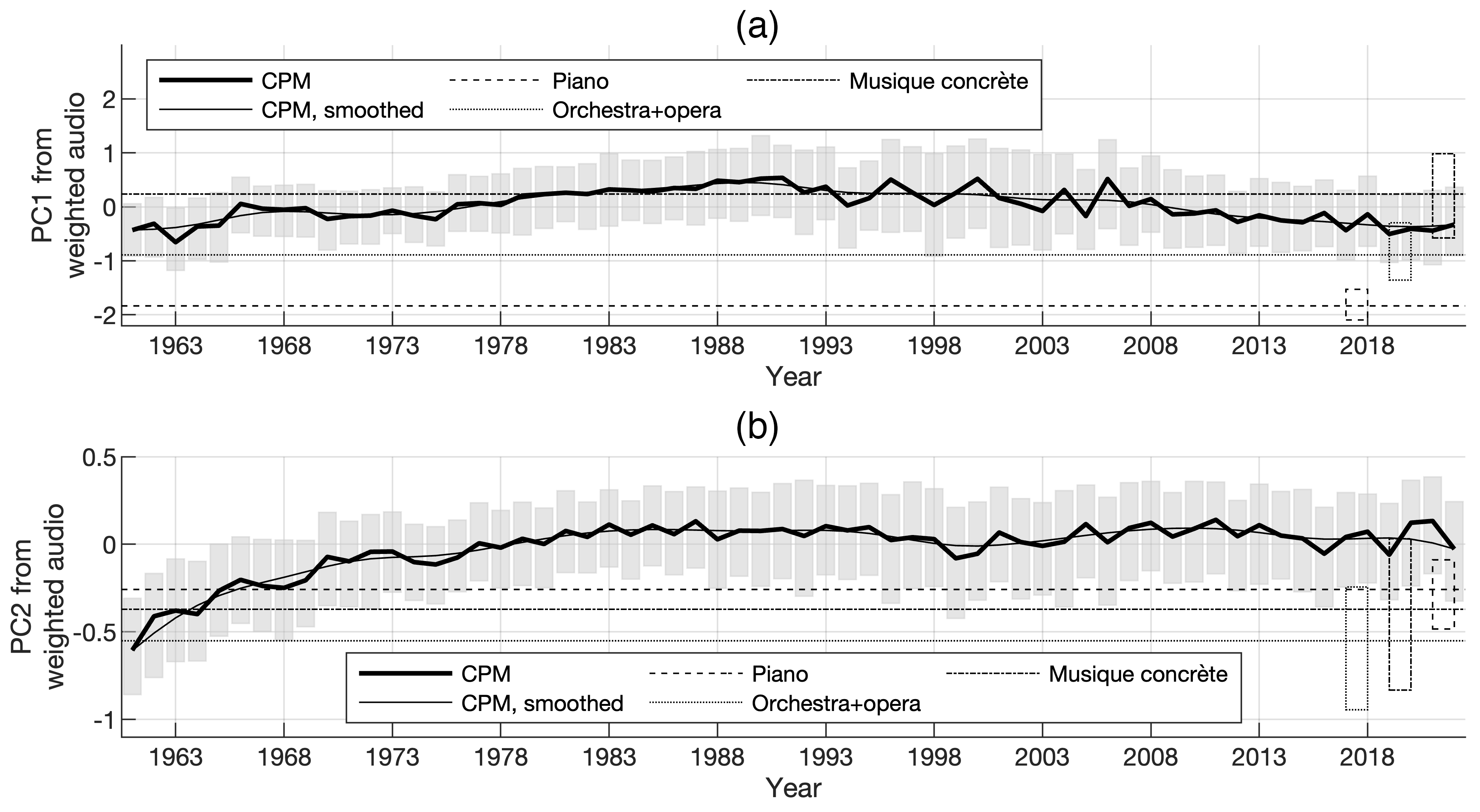}
  \caption{Measures of (a) PC1 and (b) PC2 from weighted audio for the four datasets.}
\label{fig:PCAs1D}
\end{figure}

\newpage

\subsection{High inharmonicity---to what extent?}\label{subsec:highinharm}

\begin{figure}[htbp]
  \centering
  \includegraphics[width=1\columnwidth]{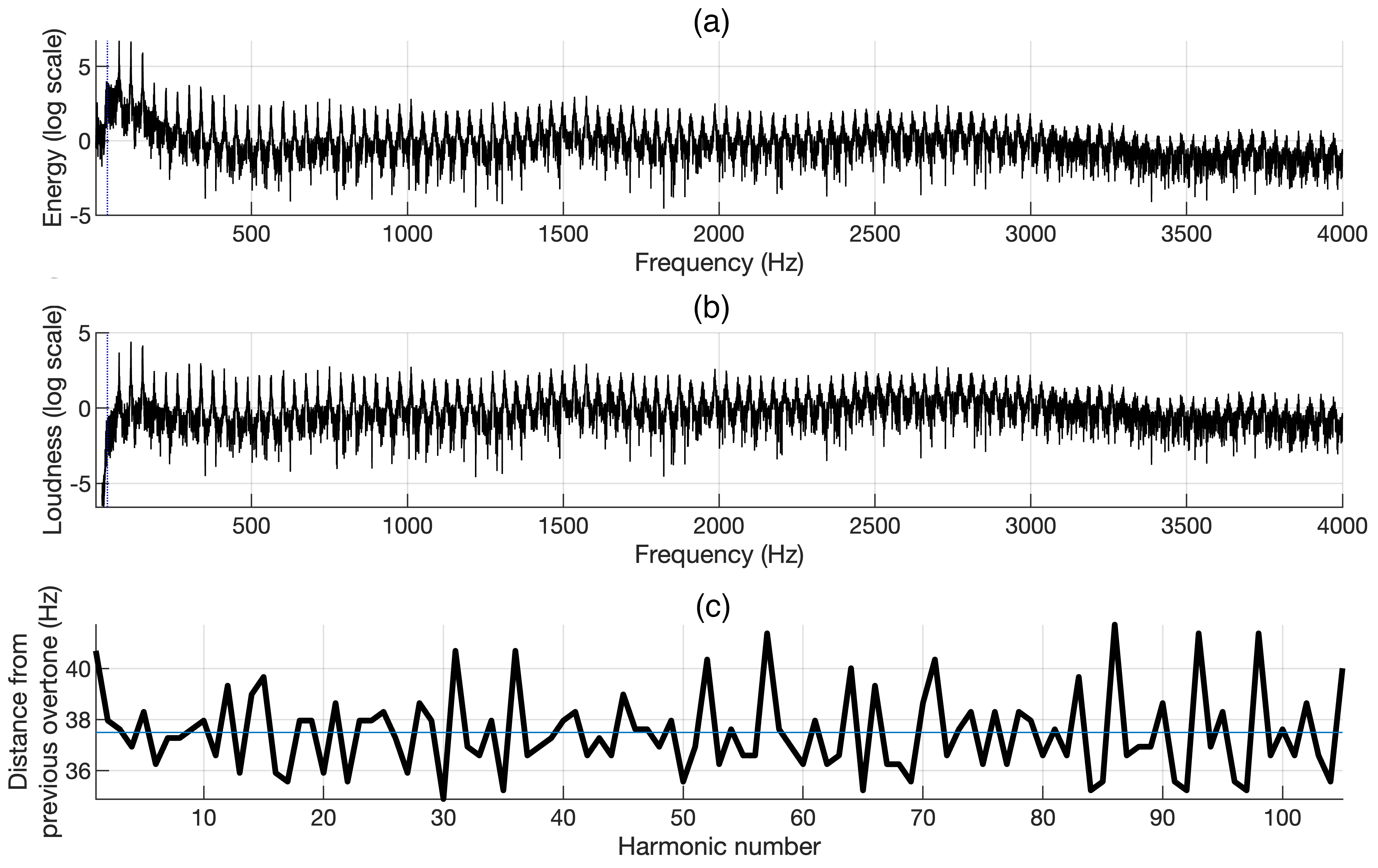}
  \caption{Excerpt from My Bloody Valentine, `Sometimes', from the album `Loveless'. The excerpt's timing is 5'06 to 5'19. (a) Power spectrum. (b) Power spectrum weighted with ISO226:2003, 50 phon. (c) Frequency difference between consecutive partials. The mean frequency difference (37.5Hz) is drawn in the top and middle diagram as the blue dotted line.}
\label{fig:Sometimes}
\end{figure}

Given that we perform the analysis on final stereo tracks, it is difficult to verify this conclusion in the general case. However, it is possible to analyse a particular example to understand one way to reach high PC2 values. The song `Sometimes' by My Bloody Valentine features particularly high PC2 values. At the end of the song, a solo guitar chord corresponds to even higher PC2 values \citep{mybloodyvalentine1991sometimes}. We isolate this part and evaluate its power spectrum. Figure~\ref{fig:Sometimes}, top, shows that the part is a quasi-harmonic 37.5Hz complex tone with its fundamental missing. Figure~\ref{fig:Sometimes}, middle, shows that the relative loudness of the overtones is high, which explains why this single complex tone was initially perceived as a chord: we hear some of the overtones as independent notes. Figure~\ref{fig:Sometimes}, bottom, shows the frequency difference between consecutive partials. The difference is not constant, which makes the complex tone strongly inharmonic.

Such properties of the signal are not specific to this particular song. For instance, we can witness a similar tone architecture in the keyboard part at the end of Alt-J's `Hunger Of The Pine' \citep{altj2014hunger}---albeit with fewer inharmonic overtones. It is also worth noting that `Loveless' has been highly influential, being an inspiration to artists such as The Verve, Oasis, Deerhunter, M83, DIIV, Deafheaven and Coldplay \citep{hudson2021loveless}. It is rated by BestEverAlbums.com as the second best album of 1991, second only to the extremely successful `Nevermind' by Nirvana.

\citet{schneider2009perception} studied inharmonicity in bells from the belfry of Brugge, noting that the SPINET pitch detection algorithm \citep{cohen1995spectral} fails to yield interpretable results with single bell samples. They also observed that the use of bells `hampers [human] detection of the pitches', complicating voice following and understanding of tonal functions. This raises the question of whether inharmonicity in CPM similarly obstructs algorithmic pitch detection, voice tracking, and tonal function comprehension.

\begin{figure}[htbp]
  \centering
  \includegraphics[width=1\columnwidth]{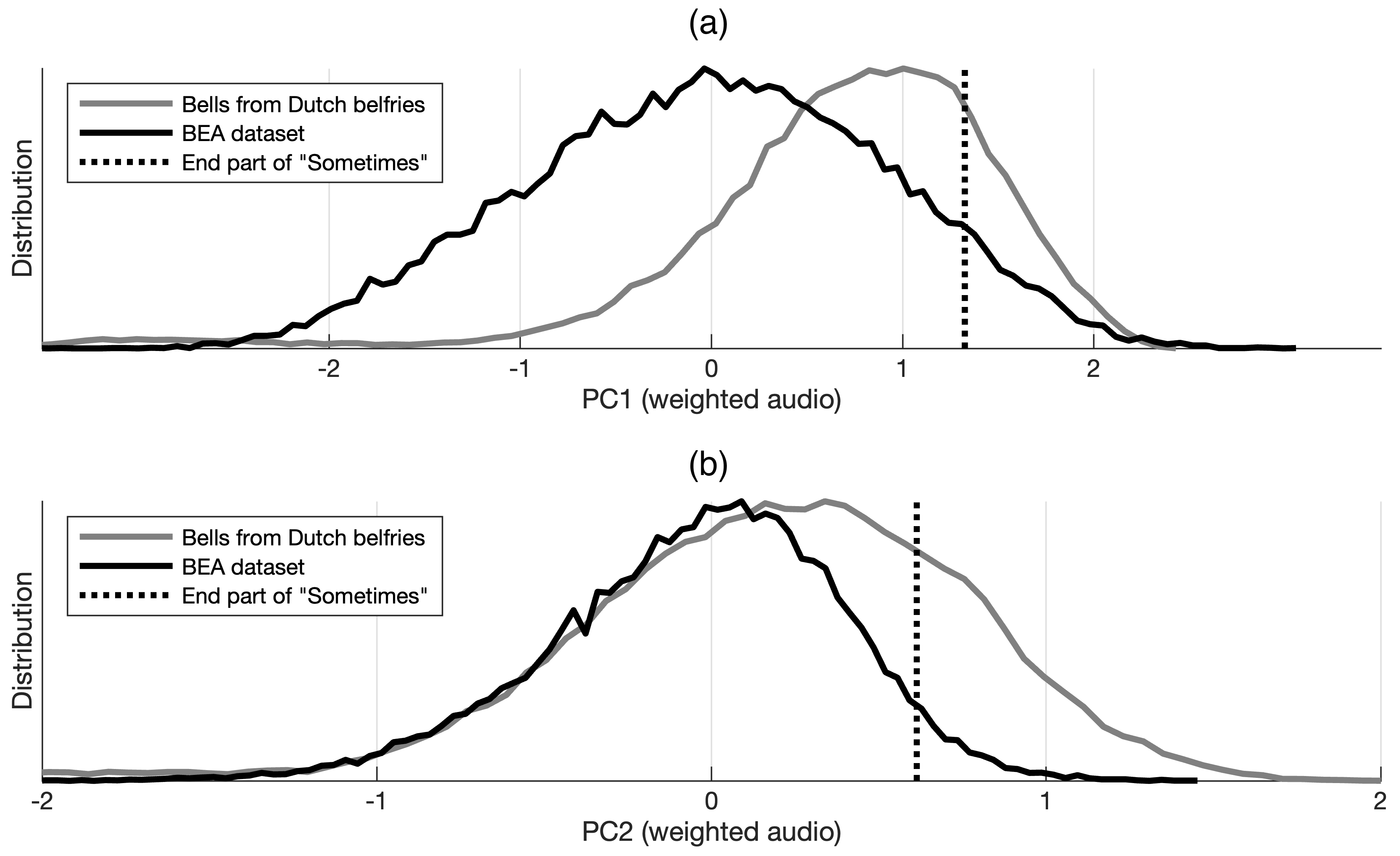}
  \caption{Comparison of normalised (a) PC1 and (b) PC2 values from weighted audio: bells, BEA dataset, and end part of My Bloody Valentine's `Sometimes'.}
\label{fig:bells}
\end{figure}

We compared PC1 and PC2 values for bells from Dutch belfries\footnote{\url{https://revivethis.org/sample-pack-bells/}} with those for the BEA dataset. Figure~\ref{fig:bells} shows the results, indicating that bells generally exhibit higher PC1 and PC2 values than popular music. However, the values for the complex tone in Figure~\ref{fig:Sometimes} (My Bloody Valentine, `Sometimes') exceed the median bell values in both PC1 and PC2. This complex tone reflects a higher total noise and inharmonicity (PC1) and a higher proportion of inharmonicity from intervals between partials (PC2) than most bell sounds. It suggests that, in CPM, some instruments (in this case, the electric guitar, which is not a marginal instrument) may introduce enough inharmonic relations between partials to hinder human and algorithmic pitch recognition. 

\section{Conclusions}

The initial question, as introduced in Section~\ref{sec:introduction} and elaborated upon in Section~\ref{subsec:extent}, can be phrased as follows: given that recent popular music, unlike Western classical music, is not subject to the constraint of (almost exclusively) using acoustic resonators (as employed in acoustic instruments), to what extent do the two music categories differ in terms of properties that can be related to the use of acoustic resonators? The short answer to this question is that recent popular music is both noisier and more inharmonic than Western classical music (Section~\ref{sec:results}, Figures \ref{fig:2Dfourdatasets} and \ref{fig:2Dfourdatasetsweighted}). It is also more inharmonic than \textit{musique concr\`ete} and slightly less noisy.

We now summarise the methods on which the study is based.

\vspace{1cm}

\begin{enumerate}

    \item We directly derived the HR-inharmonicity feature (Section~\ref{sec:HRinharmonicity}, detailed in Appendix~\ref{sec:HRinterpretation}) from the MPEG-7's \textit{HarmonicRatio} feature, which in turn has been used to measure pitch strength.
    
    \item We used the noisiness feature (Section~\ref{sec:noisinessintro}, detailed in Appendix~\ref{sec:noisiness}), derived from the MPEG-7's \textit{AudioFlatnessType} feature, primarily characterised as being the opposite of peak prominence. Noisiness denotes a locally `flat' spectrum.

     \item We argued that, as the human ear is not equally sensitive to all frequencies, weighting the audio using equal loudness contours prior to feature evaluation can provide results that better reflect what a human listener actually hears (Section~\ref{sec:audioweightingintro}, detailed in Appendix~\ref{sec:weighting}). Measures showed that the weighting increases the proximity of perceptually similar tracks and reduces the number of outliers. 

	\item As HR-inharmonicity and noisiness are not independent of each other (Section~\ref{sec:inharmintro}, detailed in Appendix~\ref{subsubsec:HRandnoise}), we used PCA to define two new features, PC1 and PC2. PC2 is particularly significant, as it measures the proportion of the total inharmonicity that is due to inharmonic relations between partials (as opposed to noise). It follows that it also measures the relative effect on pitch strength of inharmonic relations between partials and noise.

\end{enumerate}

We now summarise the main findings of the study.

\begin{enumerate}

	 \item We showed that HR-inharmonicity of popular music is comparable to that of \textit{musique concr\`ete} (Section~\ref{subsec:HRmeasures}). At the beginning of the 1960s, it was similar to that of Western orchestral music, and it then increased to reach a peak in the mid-1980s. In Appendix~\ref{sec:HRinterpretation}, we show that high HR-inharmonicity values, and, therefore, low pitch strength, may derive from noise and/or interaction between inharmonically related partials within and between complex tones. In the latter case, the degree of inharmonicity depends on both the proximity and energy of interacting partials. We also showed that higher HR-inharmonicity values prompt faster acoustic beating.

	 \item We showed that noisiness of popular music is comparable to that of Western orchestral music and lower than that of \textit{musique concr\`ete} (Section~\ref{peakprominencemeasures}). It reached a maximum during the mid-1980s.

	\item  In regards to the frequencies to which humans are the most sensitive, popular music is generally slightly noisier and considerably more HR-inharmonic (lesser pitch strength) than orchestral music (Section~\ref{sec:alldatasets}). The total amount of HR-inharmonicity and noisiness combined (PC1) is greater on average in popular music than in Western classical music. 

	\item The proportion of HR-inharmonicity that is due to inharmonic relations between partials for popular music at the beginning of the 1960s was comparable to that of orchestral music (Section~\ref{sec:alldatasets}). It then increased rapidly until ca. 1972. Later evolutions are slower. It was therefore suggested that popular music released after the evolution forms a category of its own, which we refer to as `Contemporary Popular Music' (CPM).  The density of inharmonic relations between partials within CPM is, on average, significantly higher than that of music from other categories. 

\end{enumerate}

We linked feature values with examples drawn from the popular music dataset (Sections \ref{subsec:selectedartists} and \ref{subsec:causes}) and observed the following:

\begin{enumerate}

   \item High noisiness values appear to be easy to link with the audio content. Noise may be heard in drums, vocal plosives, distortions, and `real-life' noisy ambiances. 

    \item HR-inharmonicity and PC2 are more difficult to pinpoint. We hypothesise that high HR-inharmonicity and PC2 values may derive from the use of `tonal' sounds coupled with intensive use of the studio. Such use of the studio may be specific to CPM, as \textit{musique concr\`ete}, another category of studio-based music, is both noisier and less HR-inharmonic.

\end{enumerate}

We were able to draw other conclusions that are not direct answers to the initial question but may remain useful in the context of the domain of music information retrieval:

\begin{enumerate}

        \item While high inharmonicity values may originate from both the intervals between complex tones and the inharmonicity of individual complex tones, increasing the inharmonicity of individual complex tones reduces the relative influence on inharmonicity of the intervals between complex tones (Appendix~\ref{sec:respective}).

    \item According to \citet[p. 20]{peeters2004large}, `Spectral Flatness is a measure of the noisiness (flat, decorrelation)/sinusoidality of a spectrum'. According to \citep[p.~26]{mpeg7audio}, `[t]his descriptor expresses the deviation of the signal’s power spectrum over frequency from a flat shape (corresponding to a noise-like or an impulse-like signal). A high deviation from a flat shape may indicate the presence of tonal components'. According to the Essentia documentation,\footnote{\url{https://essentia.upf.edu/reference/streaming/_Flatness.html}} spectral flatness is `a measure of how noise-like a sound is, as opposed to being tone-like'. We demonstrated that Spectral Flatness fails to give high noisiness values for pink noise, undermining its purpose. We suggested filtering out the overall spectral envelope before evaluating spectral flatness (Appendix~\ref{sec:noisiness}).

    \item In terms of noisiness and inharmonicity, we found no evidence of the `homogenization of the timbral palette' reported by \citet{serra2012measuring}. On the contrary, the evolution of popular music towards CPM is accompanied by a significant {\em increase\/} in the diversity of noisiness and inharmonicity values, which evolves to be greater than that of orchestral music (Section~\ref{sec:diachronic}).
    
        \item The output of feature-based studies may be made more reliable by attenuating the influence of audio frequencies human listeners are less sensitive to (see Appendix~\ref{sec:weighting}). In particular, the weighting function gives relatively much less weight to the lower frequencies. This suggests that it may be useful to weigh the initial signal using models such as ISO226:2003 \citep{iso2262003}.

\end{enumerate}

Finally, in Section~\ref{subsec:highinharm}, we compared the HR-inharmonicity and PC2 values in popular music with those for church bells. Although church bells generally involve more inharmonic relations between partials than popular music, we showed that there are instances in CPM of sounds in which there are more inharmonic relations between partials than most church bells. 
It has been suggested in the literature that inharmonicity in bells hampers both automatic pitch detection and human perception of pitch. We therefore expect similar difficulties to arise when detecting or perceiving pitch in passages from CPM songs where the values of HR-inharmonicity and PC2 are high. In such cases, transcribing the music into notes in staff notation may make little sense.
Using the terms introduced in Sections~\ref{subsec:resources} and \ref{subsec:extent}, idioms of popular music may not always involve musical notes, even when the elements on which the idioms are based involve waveforms that feature a strongly periodic behaviour and convey an impression of pitch. This is a topic we intend to explore further in future studies.

\appendix

\section{HR-inharmonicity and Noisiness}\label{sec:appendix1}

\subsection{A study of \textit{HarmonicRatio}}\label{sec:HRinterpretation}

In this section, we examine the \textit{HarmonicRatio} feature mentioned in Section~\ref{sec:HRinharmonicity}. Though a full explanation of `harmonicity', `inharmonicity', and the \textit{HarmonicRatio} feature is beyond this paper's scope, an intuitive grasp of the feature's behaviour is needed to interpret the subsequent \textit{HarmonicRatio} measures. 

\subsubsection{Conventions}

We use the following terminology:

\begin{enumerate}
    \item The term \emph{partial} refers to the spectral representation of a single periodic sine wave.
    \item A \emph{harmonic complex tone} denotes a group of partials that are integer multiples of a fundamental frequency.
    \item An \emph{inharmonic complex tone} refers to a tone (perceived to have a pitch) consisting of partials that are near-integer multiples of a fundamental frequency. Examples are given by \citet{fletcher1962quality} and \citet[p.~9]{rasch1982perception}.
    \item In harmonic or inharmonic complex tones, an \emph{overtone} is any partial except the one corresponding to the fundamental frequency.
    \item A {\em harmonic\/} is any sinusoidal component of a harmonic complex tone, with the $n$\textsuperscript{th} harmonic having a frequency $n$ times that of the fundamental (the first harmonic).
\end{enumerate}

In this section, harmonic complex tones are generated using the model from \citet{mauch2010approximate}, originally introduced by \citet{gomez2006tonal}, where the $k$\textsuperscript{th} partial, $a(k)$, is assigned an amplitude of $s^{k-1}$ for a constant $s < 1$. Unless specified otherwise, we use 10 harmonics with a 220Hz fundamental (A3) and set $s=0.8$.

\subsubsection{One complex tone}\label{subsubsec:onecomplextone}

\begin{figure}[htbp]
  \centering
  \includegraphics[width=1\columnwidth]{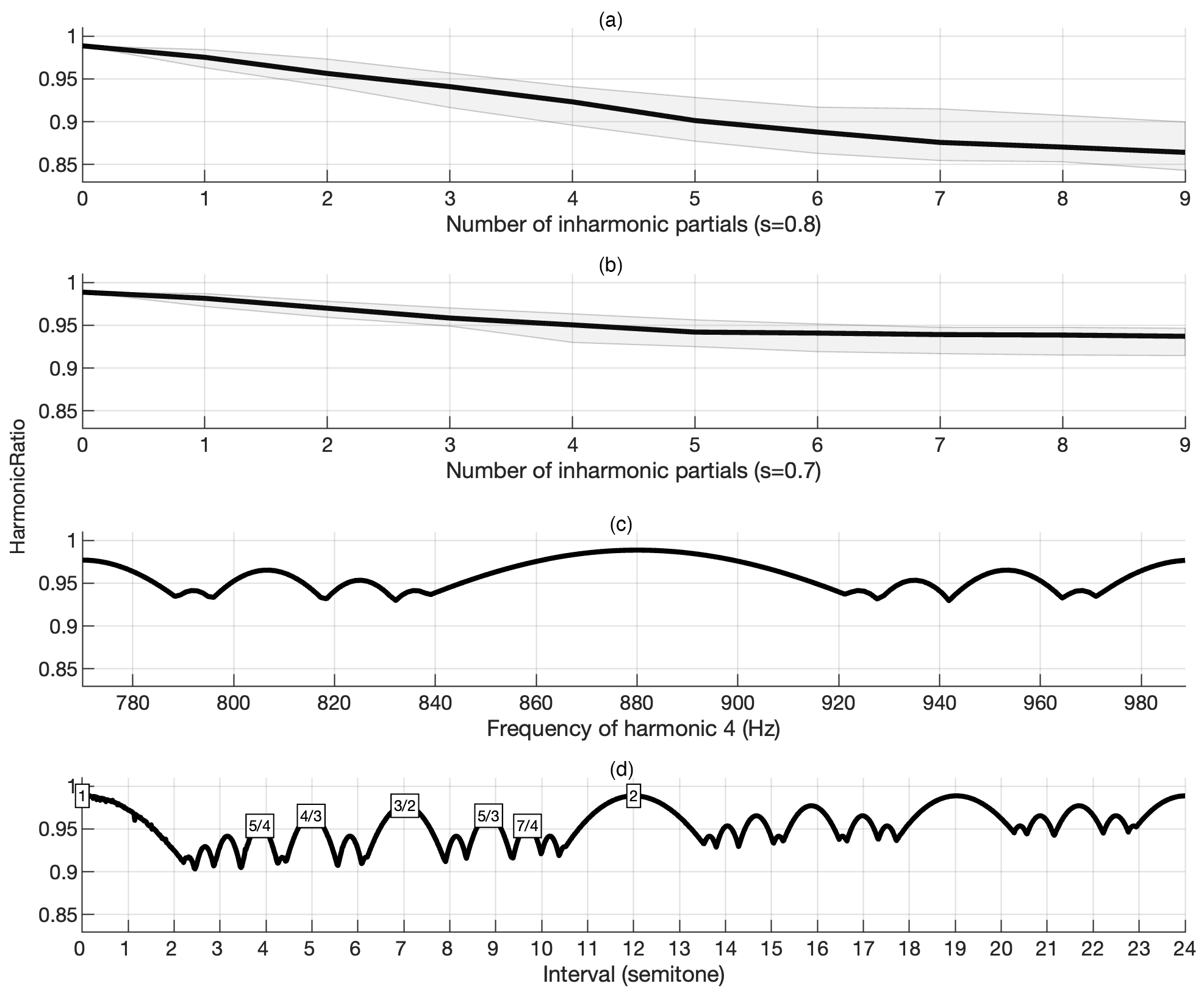}
  \caption{(a) \textit{HarmonicRatio} resulting from an increasing number of inharmonic partials in an initially harmonic complex tone ($f_0=200$Hz, 10 harmonics). The inharmonicity of the partials is random, the representation summarises observations from 50 tones, the black line denotes the median over the 50 observations, and the gray area the 25\textsuperscript{th} and 75\textsuperscript{th} percentiles. (b) Same as top graph, but with  harmonics of lower energy. (c) \textit{HarmonicRatio} resulting from the frequency shift of a single partial. (d) \textit{HarmonicRatio} resulting from the superposition of two sine waves depending on the interval between them. The text boxes show the frequency ratios of the `purer' intervals inside the first octave. The $y$-axis scale is identical for all graphs in order to facilitate comparison.} 
\label{fig:influenceofnumberofofinharmonicpartials}
\end{figure}

We begin with the case of a single isolated complex tone. Figure~\ref{fig:influenceofnumberofofinharmonicpartials} shows the evolution of \textit{HarmonicRatio} values in four cases:

\begin{enumerate}
    \item Increasing the number of inharmonic partials in an initially harmonic tone.
    \item Increasing the number of inharmonic partials in an initially harmonic tone, with harmonics having less energy than in case 1 ($s=0.7$).
    \item Frequency shift of a single partial (here, partial 4) in an initially harmonic tone.
    \item Superposition of two sine waves of equal power. The first sine wave has a frequency of 220Hz, while the second ranges from 220Hz to 880Hz.
\end{enumerate}

The results are shown in Figure~\ref{fig:influenceofnumberofofinharmonicpartials}, from which we can derive the
following observations:

\begin{enumerate}
\item Figure~\ref{fig:influenceofnumberofofinharmonicpartials} (a) confirms the MPEG-7 specification's interpretation \citep{mpeg7audio}, showing that increasing inharmonic components in an initially harmonic complex tone decreases the \textit{HarmonicRatio} value.
\item Figure~\ref{fig:influenceofnumberofofinharmonicpartials} (b) shows that \textit{HarmonicRatio} depends on the energy of the inharmonic components. As these components grow more intense, the \textit{HarmonicRatio} value drops.
\item Figure~\ref{fig:influenceofnumberofofinharmonicpartials} (c) shows that when one partial is inharmonic, \textit{HarmonicRatio} depends on the partial's frequency.
\item Figure~\ref{fig:influenceofnumberofofinharmonicpartials} (d) shows the influence of intervals between two partials on \textit{HarmonicRatio}. (i) Simpler intervals, i.e., fractions with smaller integers \citep{vos1985thresholds}, are local maxima. The denominator influences the interval's \textit{HarmonicRatio}. (ii) Wider intervals yield higher \textit{HarmonicRatio} values. These findings recall sensory dissonance \citep[p.46, p.85]{sethares2005tuning}, and what \citet{masina2022dyad} refer to as `roughness', following \citet{hutchinson1978acoustic} and \citet{plomp1965tonal}.
\end{enumerate}

 To summarise, given one complex tone, low \textit{HarmonicRatio} values may be the combined result of the following:

\begin{enumerate}
    \item
    a high proportion of intervals between partials that are not close to `pure' intervals; and
    \item the intervals that are far from `pure' intervals involve partials with high energy values.
\end{enumerate}

\subsubsection{Several complex tones}\label{subsubsec:severalcomplextones}

The factors leading to low \textit{HarmonicRatio} values in single complex tones also apply when several complex tones are superposed. One difference is that the distribution of the complex tones' fundamentals often reflects scales or tuning systems (e.g., diatonic scale or 12-tone equal temperament). We explore the \textit{HarmonicRatio} values arising from the superposition of complex tones with fundamentals conforming to different scales. Starting with one harmonic complex tone, we add others with fundamentals ranging from 220Hz to 2200Hz. We compare the \textit{HarmonicRatio} measures in four cases:

\begin{enumerate}
    \item the fundamental frequencies can take on any value;
    \item their values are restricted to a chromatic scale using equal temperament;
    \item their values are restricted to major triads using equal temperament; and
    \item their values are restricted to major triads using the pure intervals shown in Figure~\ref{fig:influenceofnumberofofinharmonicpartials}.
\end{enumerate}

\begin{figure}[htbp]
  \centering
  \includegraphics[width=1\columnwidth]{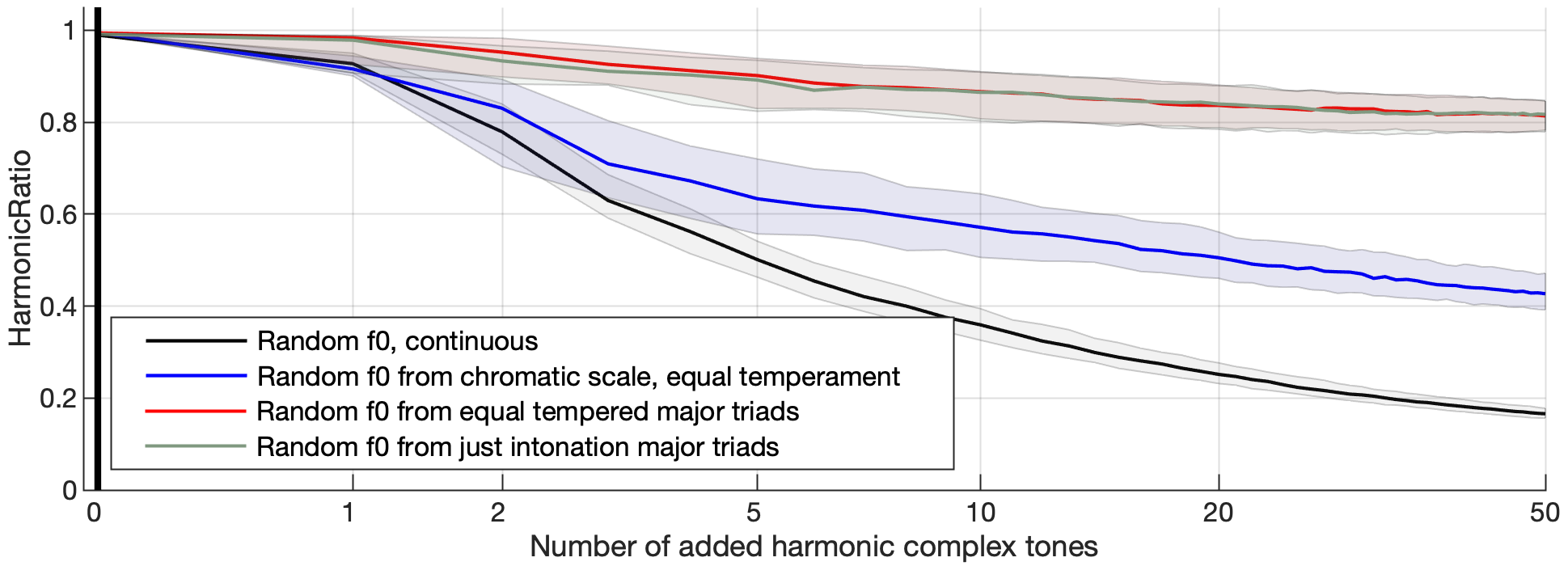}
  \caption{\textit{HarmonicRatio} resulting from the sum of harmonic complex tones, depending on the distribution of fundamental frequencies. As the fundamental frequencies are randomly sampled from their respective distribution, the measures are performed 500 times. The solid lines represent the median, and the areas are bounded by the 25\textsuperscript{th} and 75\textsuperscript{th} percentiles.}
\label{fig:complex_tones_to_inharmonicity}
\end{figure}

 The results are shown in Figure~\ref{fig:complex_tones_to_inharmonicity}, from which we can observe the following:
 
\begin{enumerate}
    \item A higher number of harmonic tones leads to lower \textit{HarmonicRatio} values. In other words, more voices in the music increase inharmonicity.
    \item \textit{HarmonicRatio} values are affected by the scales used in the music. \textit{HarmonicRatio} is lowest for a continuous scale (case 1). It increases for the twelve-tone chromatic scale (case 2), rises further with major triads (case 3), and remains similar when pure intervals replace equal-tempered intervals in major triads (case 4).
    \item Very low \textit{HarmonicRatio} values from fundamentals not following a discrete scale suggest that continuous frequency contours (e.g., human voices) may lower \textit{HarmonicRatio} values.
\end{enumerate}

Two factors influencing \textit{HarmonicRatio} values for multiple complex tones stem from the observations in Appendix~\ref{subsubsec:onecomplextone} for a single complex tone:

\begin{enumerate}
    \item{\em Proximity of $f_0$s\/}. Closer partials tend to lower \textit{HarmonicRatio} values. Thus, nearby fundamental frequencies (e.g., in near-coincident musical voices) lead to lower \textit{HarmonicRatio} values.
    \item{\em Voices of equal energies\/}. When all partials are inharmonic (e.g., piano sounds \citep{fletcher1962quality}), the lowest \textit{HarmonicRatio} values occur when partials have equal energy. If the fundamentals are close in frequency, the lowest \textit{HarmonicRatio} values arise when they share equal energy.
\end{enumerate}

To summarise, given several complex tones, low \textit{HarmonicRatio} values may be the combined result of:

\begin{enumerate}
    \item a high number of simultaneous voices;
    \item voices being close together in frequency;
    \item voices having the same level;
    \item musical intervals not being `pure'; and
    \item a continuous distribution of $f_0$ values.
\end{enumerate}

\subsubsection{Respective contribution of fundamental and overtones}\label{sec:respective}

Figure~\ref{fig:two_tones_decomposed} (a) and (b) show the \textit{HarmonicRatio} values from the pairwise superposition of components of two harmonic complex tones. The tones in Figure~\ref{fig:two_tones_decomposed} (a) and (c) have fundamentals 3 semitones apart (an equal-tempered minor third), while those in Figure~\ref{fig:two_tones_decomposed} (b) and (d) have fundamentals 5 semitones apart (an equal-tempered perfect fourth).

As shown in Figure~\ref{fig:influenceofnumberofofinharmonicpartials} (d), for two sine waves, the equal-tempered minor third yields a lower \textit{HarmonicRatio} than the perfect fourth. In the case of complex tones, we can distinguish between harmonic and inharmonic complex tones:

 \begin{enumerate}
     \item {\bfseries Two {\em harmonic\/} complex tones (Figure~\ref{fig:two_tones_decomposed} (a) and (b))$\quad$} For both the minor third and the perfect fourth, \textit{HarmonicRatio} values from the superposition of overtones within a single harmonic complex tone are higher than those from the superposition of overtones between different harmonic complex tones. This applies to all intervals except the octave, where all harmonics of the upper tone coincide with those of the lower tone, making the \textit{HarmonicRatio} values from interactions within one tone the same as those between different tones.

\begin{figure}[H]
  \centering
  \includegraphics[width=1\columnwidth]{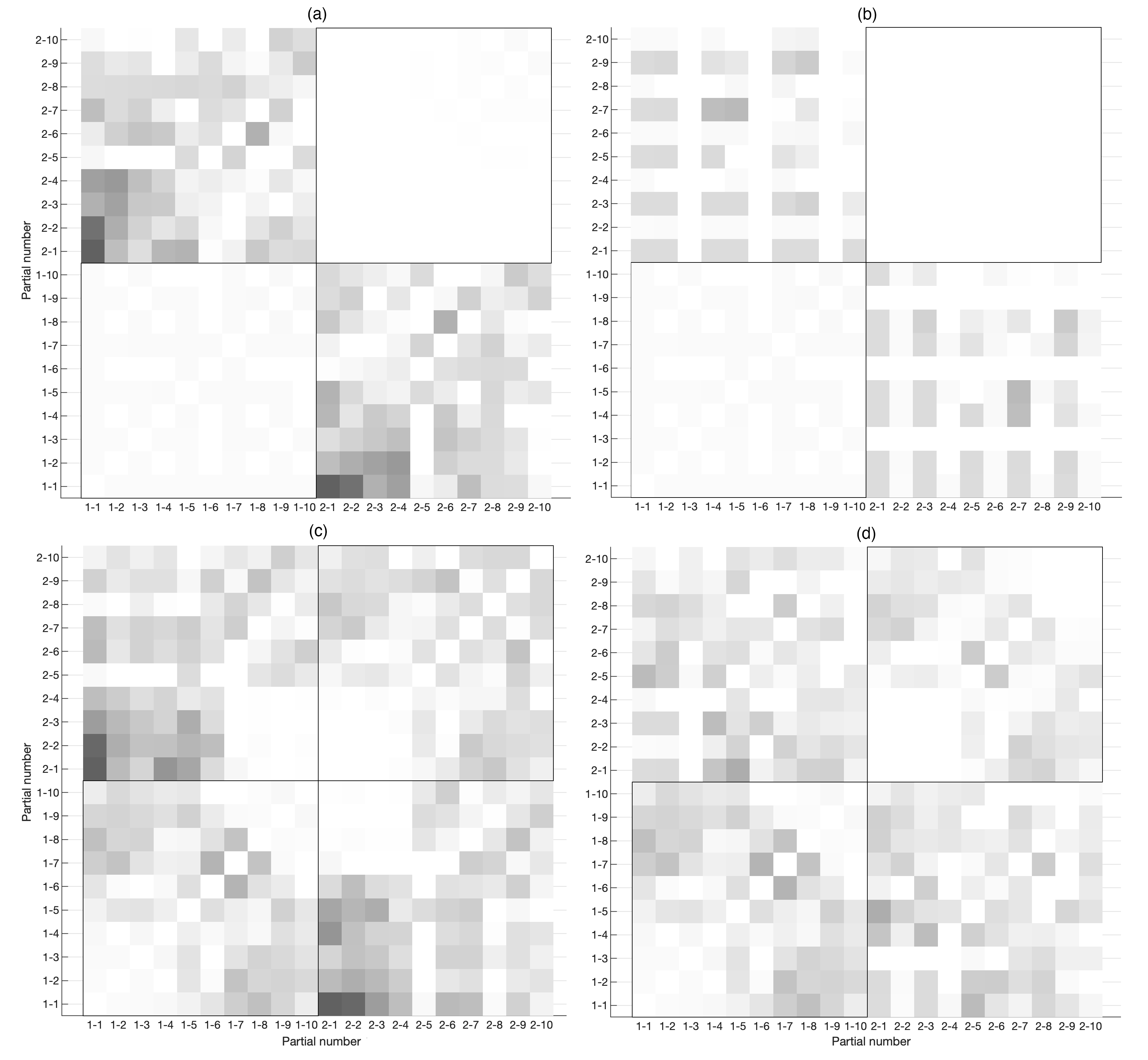}
  \caption{(a) and (b) show \textit{HarmonicRatio} values derived from the pairwise combination of elements in two harmonic tones. Clearer shades correspond to higher values, and darker shades to lower values. The interval between the two tones is 3 half-tones in (a) and 5 half-tones in (b). Partial numbers are written as `tone number-partial number' (e.g., `2-3' corresponds to the third partial of the second tone). (c) and (d) show \textit{HarmonicRatio} values derived from the pairwise combination of elements in two inharmonic tones (same fundamentals as in (a) and (b), respectively). The inharmonicity coefficient in (c) and (d) is 0.002 as defined by \citet{fletcher1962quality}.}
\label{fig:two_tones_decomposed}
\end{figure}

     \item {\bfseries Two {\em inharmonic\/} complex tones (Figure~\ref{fig:two_tones_decomposed} (c) and (d))$\quad$} \textit{HarmonicRatio} values from the superposition of partials within the tones are generally lower than for harmonic complex tones. For the five-semitone interval, they are lower than those from the superposition of the fundamentals. This observation would hold for all intervals with a high \textit{HarmonicRatio} value.

 \end{enumerate} 

To summarise:

\begin{enumerate}
     \item Low \textit{HarmonicRatio} values may originate from both the intervals {\em between\/} complex tones and the inharmonicity of individual complex tones.
     \item Increasing the inharmonicity of individual complex tones reduces the relative influence on the inharmonicity of the intervals between complex tones.
\end{enumerate}

\subsubsection{\textit{HarmonicRatio} and acoustic beating}\label{subsec:beating}

As noted in Appendix~\ref{subsubsec:onecomplextone}, the behaviour in the bottom graph of Figure~\ref{fig:influenceofnumberofofinharmonicpartials} resembles that of `roughness'. According to \citet{masina2022dyad}, the relation between `roughness' and acoustic beats was introduced by Foder{\`a} in 1832--1837 \citep{barbieri2002nascita} and later developed by Helmholtz in the 19\textsuperscript{th} century \citep[pp. 197-211]{helmoltz1885sensations}. \citet[p. 15]{rasch1982perception} explain how acoustic beating corresponds to `roughness' when the interval between two tones is smaller than a critical band. We compare measures of acoustic beating with \textit{HarmonicRatio} values from the superposition of harmonic complex tones.

\begin{figure}[htbp]
  \centering
  \includegraphics[width=1\columnwidth]{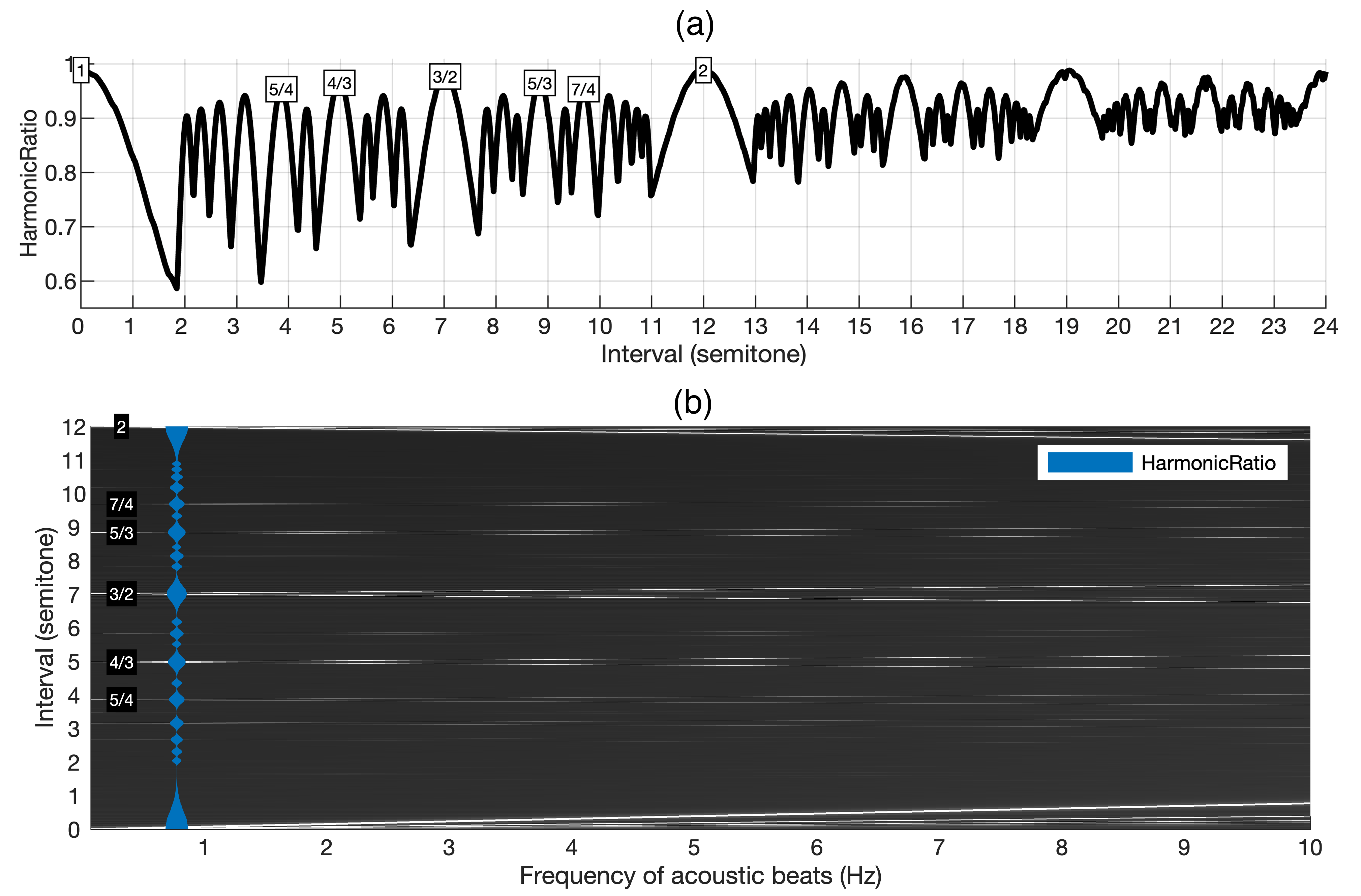}
  \caption{(a) \textit{HarmonicRatio} resulting from the superposition of two harmonic complex tones depending on the interval between them (10 harmonics, $s=0.8$) As in Figure~\ref{fig:influenceofnumberofofinharmonicpartials}, bottom, the text boxes show the frequency ratios of the `purer' intervals inside the first octave. (b) Frequency beating resulting from the superposition of two harmonic complex tones. The $y$-axis corresponds to the first part of the top graph's $x$-axis. The $x$-axis is the frequency of the signal's RMS. White values indicate a high RMS variation, in other words, a high acoustic beating amplitude. The width of the superposed blue area denotes the \textit{HarmonicRatio} values as shown in the top graph. The location of the blue area along the $x$-axis is arbitrary.}
\label{fig:acousticbeating}
\end{figure}

Figure~\ref{fig:acousticbeating} (a) follows the same process as Figure~\ref{fig:influenceofnumberofofinharmonicpartials} (d) but uses two harmonic complex tones instead of two sine waves. The $f_0$ of the first tone is 220Hz, while the second ranges from 220Hz to 880Hz. In Figure~\ref{fig:acousticbeating} (b), we superpose the \textit{HarmonicRatio} values from Figure~\ref{fig:acousticbeating} (a) with acoustic beating measures (background image). Acoustic beating is evaluated by computing the power spectrum of the envelope (0.05s windows) of the sum of the two tones. High power spectrum values indicate the presence of acoustic beating at that frequency. A higher number of harmonics (40) and a higher $s$ factor (0.95) are used to enhance the visibility of the white lines from acoustic beating. From these results, we make the following observations:

 \newpage

\begin{enumerate}
    \item Pure intervals correspond to infinitely slow acoustic beating (i.e., no beating). Greater deviation from pure intervals leads to faster acoustic beating. Thus, increased `roughness' may correspond to faster acoustic beats.
    \item Local maxima of \textit{HarmonicRatio} values occur when acoustic beating is slowest. Therefore, high \textit{HarmonicRatio} values are associated with low `roughness' and slow acoustic beating.
\end{enumerate}

From this perspective, results involving \textit{HarmonicRatio} (such as in Section~\ref{sec:results}) may also apply to the period of acoustic beating, i.e., the `roughness' or acoustical `grit'.

\subsubsection{\textit{HarmonicRatio} and noise}\label{subsubsec:HRandnoise}

The tones used to produce the results in Figures \ref{fig:influenceofnumberofofinharmonicpartials}--\ref{fig:acousticbeating} were composed of a finite number of sinusoidal components and did not contain noise. As mentioned earlier with reference to \citet{schneider2009perception}, inharmonicity can also arise from noise. In this case, low \textit{HarmonicRatio} values can be understood as resulting from inharmonic relations between the (infinite set of notional) sine waves constituting the noise, whose frequencies can differ infinitesimally.

From the perspective of discrete spectral transforms, an infinite-length signal can be identified as noise-free when spectral bins with non-zero values are separated by bins with zero values. Conversely, noise may be characterised by several contiguous bins with non-zero values. However, defining `noise' in a spectral transform is challenging for at least the following two reasons:

\begin{enumerate}
    \item Even the spectrum of a harmonic tone is not a true line spectrum due to its finite duration.
    \item A transform occurs over an audio window, and a sine wave with slight frequency changes within the window (e.g., from vibrato) will produce a frequency distribution, not a line.
\end{enumerate}

As it is difficult to distinguish noise from spectral lines in practice, we choose not to distinguish between them. We fall back on the objective information we have. Thus, from a spectral domain perspective, the signal's information is represented as values attached to spectral bins. Frequency intervals are only defined between the spectral bin center frequencies, and any information we derive from the signal originates from the values attached to the spectral bins. From this point of view, a feature representing the `degree of inharmonicity' in the signal may be conceptualised as deriving from the ensemble of pairwise combinations between all frequency bins (positions and energy values). As a result, a feature representing the `degree of inharmonicity' in the signal will not be able to distinguish between `inharmonic
sounds which have little if any relevance for music (e.g., white or pink noise)' \citep{schneider2009perception} and `coherent'  inharmonic signals, which `sound as stable and smooth as harmonic signals' \citep{deboer1956pitch}, i.e., inharmonic complex tones with a finite number of sinusoidal components.

In light of the relation between \textit{HarmonicRatio} and pitch strength mentioned at the beginning of Section~\ref{sec:HRinharmonicity}, the `first peak of the
auto-correlation function' \citep{patterson1996relative,yost1996pitch,yost1997pitch,shofner2002pitch} is unable to determine whether a lower pitch strength results from relations between partials, or from noise. We therefore need to find a method to estimate the relative extent to which inharmonicity results from the interaction between (approximately) discrete partials as opposed to noise.

\begin{figure}[htbp]
  \centering
  \includegraphics[width=1\columnwidth]{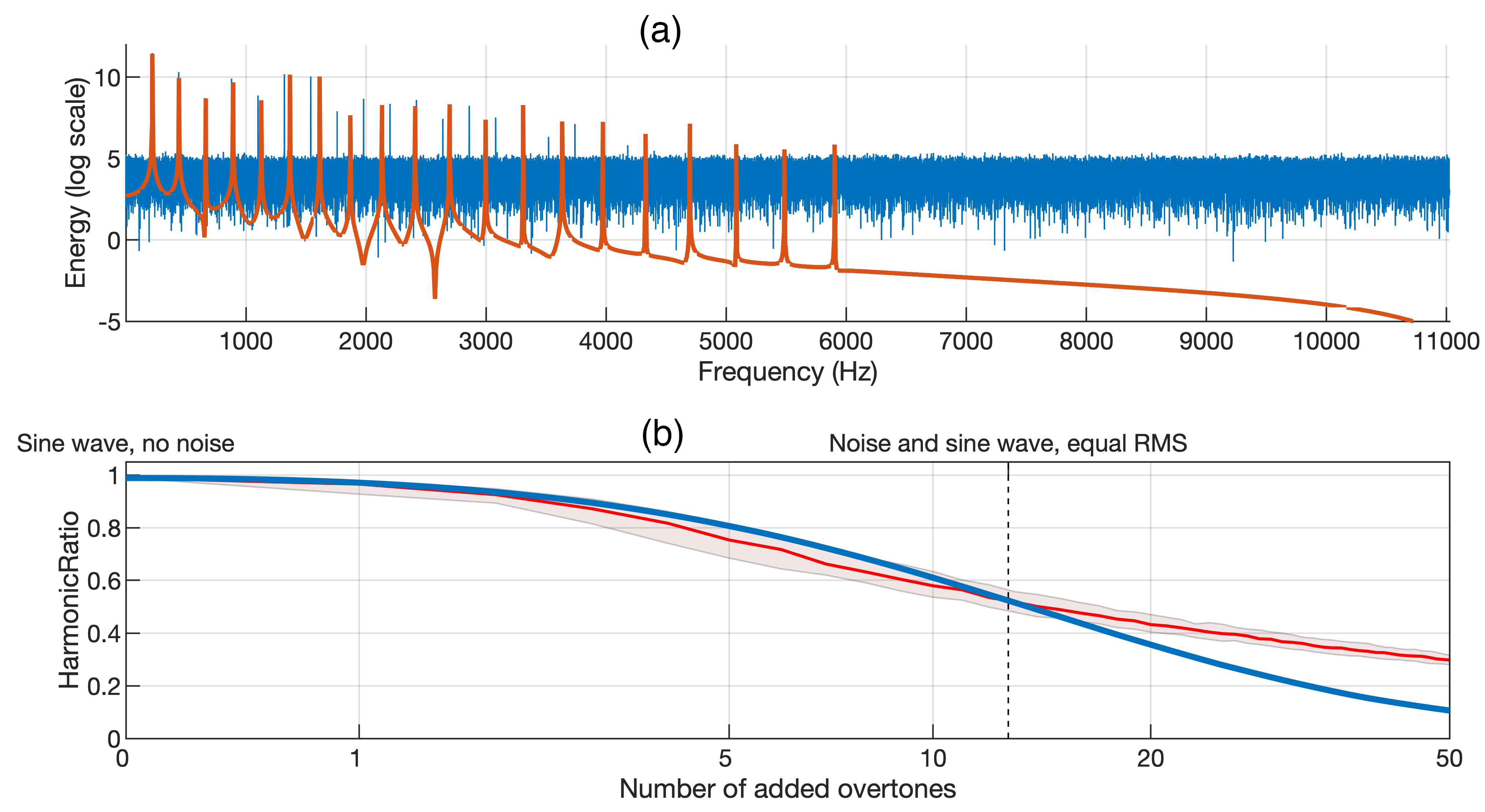}
  \caption{\textit{HarmonicRatio}, inharmonicity and noisy signals. (a) The signals corresponding to the two spectra have the same \textit{HarmonicRatio} value of 0.9558. The red signal is inharmonic, with an inharmonicity coefficient of 0.002 according to \citet{fletcher1962quality}. The blue signal is the sum of a harmonic complex tone and white noise. The white noise's RMS power is 13.5dB below that of the harmonic tone. (b) Comparative decrease of \textit{HarmonicRatio} values starting from a sine wave. The blue curve is the result of added white noise. The red curve is the result of adding sine waves of random frequencies. The frequencies being random, the experiment is repeated 200 times. The red area shows the 25\textsuperscript{th} and 75\textsuperscript{th} percentiles of the results.}
\label{fig:same_inharmonicity}
\end{figure}

Figure~\ref{fig:same_inharmonicity} (a) shows how an inharmonic complex tone and a harmonic complex tone with white noise can produce the same \textit{HarmonicRatio} value. Conversely, for a \textit{HarmonicRatio} value less than 1, the original audio may either be a non-inharmonic tone with added noise or a noiseless inharmonic complex tone.

Figure~\ref{fig:same_inharmonicity} (b) illustrates how a sum of sine waves and a sine wave with white noise can result in the same \textit{HarmonicRatio} values. The graph shows that a sine wave superposed with noise of equal RMS power is, on average, as inharmonic as a sum of 14 sine waves with the same energy and random frequencies.

\subsubsection{Inharmonicity as 1$-$\textit{HarmonicRatio}.}

The output of the \textit{HarmonicRatio} feature reflects the pairwise relations between values in all spectral bins, whether these values originate from noise or partials. High \textit{HarmonicRatio} values indicate harmonic partials without noise. Any deviation lowers the \textit{HarmonicRatio} values. This deviation can come from inharmonicity in complex tones, inharmonic relations from certain combinations of complex tones, or inharmonic relations arising from contiguous non-zero spectral bins, which may be due to noise. This summary defines an `inharmonicity' feature, termed `HR-inharmonicity', directly derived from \textit{HarmonicRatio}:

\vspace{-.5cm}

$$\mathit{HR\mbox{-}inharmonicity}=1-\mathit{HarmonicRatio}\,.$$



\subsection{`Noisiness', spectral flatness and peak prominence}\label{sec:noisiness}

In the previous section, we stated that:

\begin{enumerate}
    \item both a harmonic complex tone with noise and a noiseless inharmonic complex tone may lead to the same value of \textit{HarmonicRatio}; and that
    \item both a discrete ensemble of sine waves with no harmonic relations and white noise added to a sine wave may lead to the same value of \textit{HarmonicRatio}. 
\end{enumerate}

In order to discriminate between noiseless and noisy signals leading to the same value of \textit{HarmonicRatio}, we need a feature that measures the noisiness of the signal. To that end, we introduce a new metric derived from the \textit{AudioSpectrumFlatnessType} \citep{mpeg7audio} / `spectral flatness' \citep{peeters2004large}.

\citet[p. 20]{peeters2004large} defines spectral flatness as `a measure of the noisiness [...] / sinusoidality of a spectrum'. It is calculated as the ratio of the geometric mean to the arithmetic mean, equivalent to the Wiener entropy (WE) of the energy spectrum. A comparison between the WE and the variance of power spectra from the BEA dataset shows a high correlation (Pearson correlation = 0.9973). Thus, in this study, spectral flatness can be understood as similar to the variance of the power spectrum. Intuitively, low spectral flatness values (high WE and high variance) indicate spectral peaks, while high values (low WE and low variance, a `flat' spectrum) indicate a noise-like signal.

Spectral flatness is available in the Essentia toolbox \citep{bogdanov2013essentia} as \textit{FlatnessDB}\footnote{\url{https://essentia.upf.edu/reference/std_FlatnessDB.html}} and in the MPEG-7 standard as \textit{AudioSpectrumFlatnessType} \citep[p. 26]{mpeg7audio}. The MPEG-7 standard uses a 1/4 octave logarithmic frequency resolution to evaluate the spectrum. We prefer a higher resolution (1/4 of a semitone) to better define spectral peaks. This interval, close to a Pythagorean comma, aligns with the minimum perceivable pitch difference \citep{zarate2012pitch} or `just noticeable difference' \citep{stern2010just}. This higher resolution is feasible because the analysis uses the Fourier transform rather than the short-term Fourier transform (MPEG-7).

We calculate spectral flatness for the music datasets, pink noise, and white noise. The output is $0.998$ for white noise. For pink noise ($0.84$), the value falls between the BEA dataset ($0.83$) and the orchestra dataset ($0.86$). The difference between white and pink noise values arises from using a logarithmic frequency scale. For white noise, bands have constant frequency width, whereas for pink noise, they decrease. On a linear scale, this is reversed: spectral flatness is sensitive to the global spectral envelope. The similarity in values for pink noise and music undermines the goal of measuring noisiness. To assess peak prominence (i.e., noisiness) independently of the global envelope, we apply median filtering to the spectrum before calculating WE. The filter uses 8-bin windows (1 tone), with experiments showing comparable results for window widths between 4 and 16 bins.

When using median filtering, the spectral troughs may have values close to zero, both positive and negative. This creates a problem for WE, which uses a geometric mean numerator and cannot be calculated with negative values. One solution is to zero all negative values, but as shown by \citet[p.~20]{peeters2004large}, this always results in WE being zero. To avoid this, we scale the filtered spectrum so that its minimum value is one. However, this makes WE sensitive to gain, meaning two identical audio samples at different levels will have different WE values. To solve this, we first normalise the audio samples using RMS power to make the measure gain-robust. Lastly, to normalise the distribution and following \citeauthor{peeters2004large}' \citeyearpar[p.~20]{peeters2004large} recommendations for the `tonality measure', we express the result on a log scale. We call this feature `peak prominence'.

To summarise, the steps leading to the measure of peak prominence are: RMS normalisation, power spectrum evaluation (log frequency scale, 25-cent wide bands), median filtering (8-bin windows), Wiener Entropy, and log of the result:

\vspace{-.5cm}

$$\mathit{Peak\:prominence}=\log(\mathit{WE}(\mathit{median\mbox{-}filtered\:spectrum\:of\:RMS\mbox{-}normalised\:audio}))\,.$$

Peak prominence was designed to be the opposite of `noisiness', which leads to:

\vspace{-.5cm}

$$\mathit{Noisiness}=-\mathit{peak\:prominence}\,.$$


\section{ELC weighting}\label{sec:weighting}

This section elaborates on the difference between analyses performed using raw and weighted audio that was introduced in Section~\ref{sec:audioweightingintro}.

\begin{figure}[htbp]
  \centering
  \includegraphics[width=1\columnwidth]{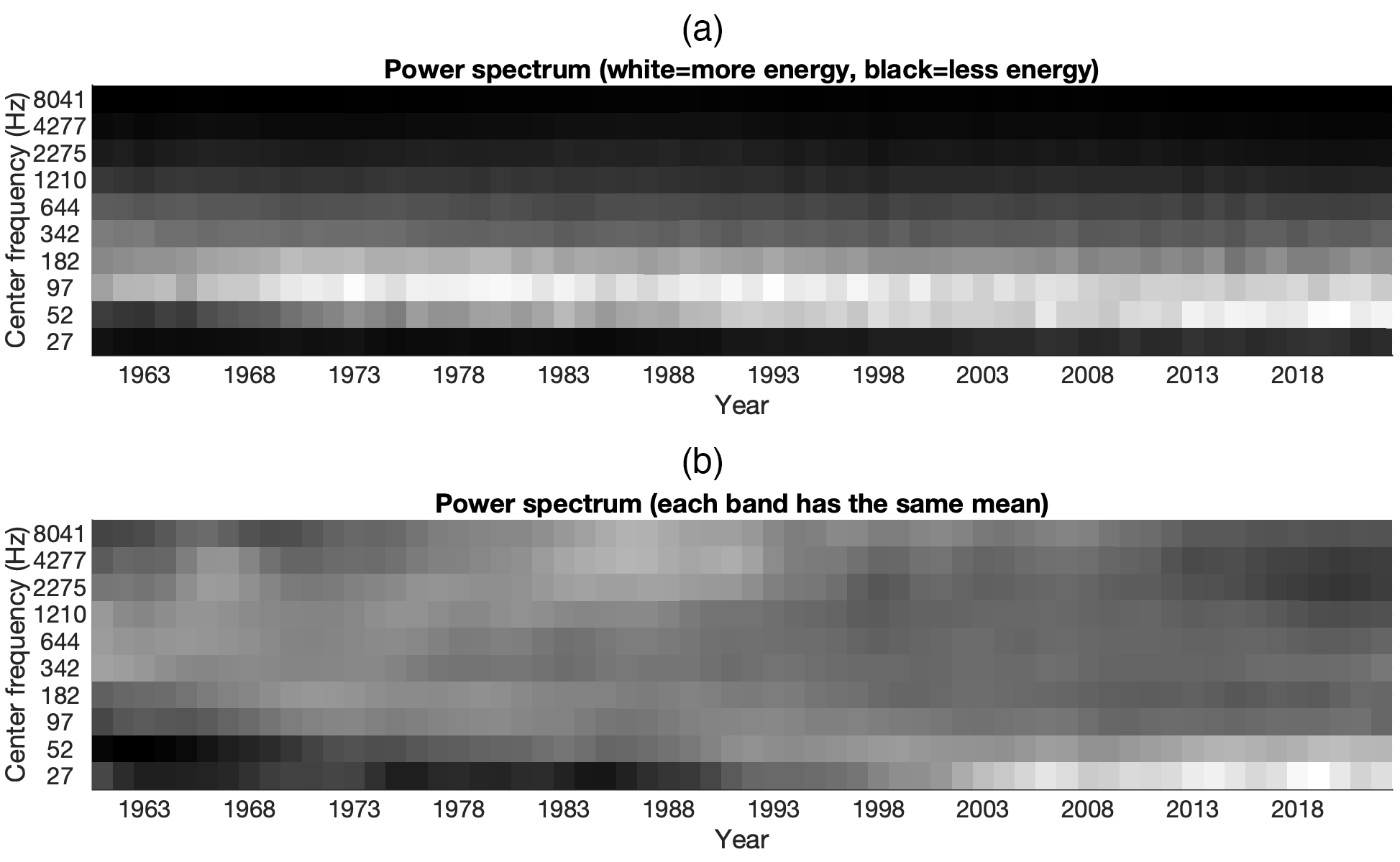}
  \caption{Power spectrum, BEA. (a) Spectrum for raw values. (b) Spectrum transformed so that values for each band have the same mean. The lower graph is blurred for better readability.}
\label{fig:allspectra10bands}
\end{figure}

Figure~\ref{fig:allspectra10bands} (a) shows the power spectrum values for the BEA dataset by year of release. Figure~\ref{fig:allspectra10bands} (b) displays the same values, normalised using each band's mean. The lower graph highlights a clear increase in power in the lowest two bands over time. This rise in bass levels is similarly documented by \citet{hove2019increased}. Over the period covered by the BEA dataset, the lowest frequencies became increasingly manipulable due to advancements in recording technology \citep{fine2008dawn}. Note the local energy peak above 2275Hz around 1986. The energy of overtones during this period is higher. As discussed in Appendices~\ref{subsubsec:onecomplextone} and \ref{subsubsec:severalcomplextones}, louder inharmonic partials lead to lower \textit{HarmonicRatio} values. This local energy maximum above 2275Hz around 1986 could thus contribute to the local peak in HR-inharmonicity at that time.

\begin{figure}[H]
  \centering
  \includegraphics[width=1\columnwidth]{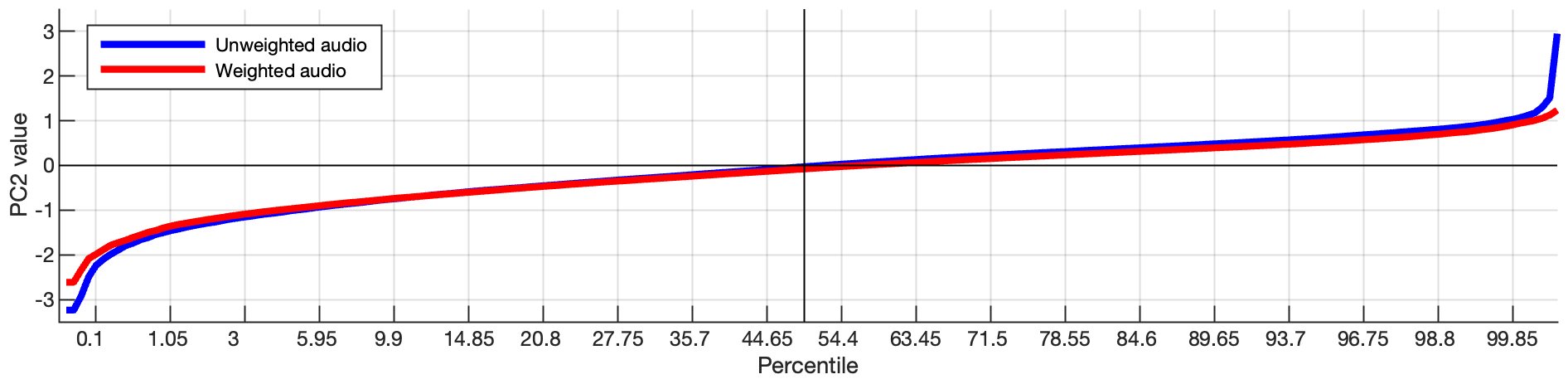}
  \caption{BEA dataset, PC2, unweighted and weighted audio, correspondence between percentiles and values.}
\label{fig:PC2_outliers}
\end{figure}

Figure~\ref{fig:2Dfourdatasetsweighted} in Section \ref{sec:HRpeakweighted} shows the PC1 and PC2 values from weighted audio for the four datasets. One key difference between Figures~\ref{fig:2Dfourdatasets} (original audio) and \ref{fig:2Dfourdatasetsweighted} (weighted audio) is that the four datasets are more clearly separated (obvious in the case of piano, orchestra, and popular music). The weighted data appears to be less long-tailed, with fewer outliers. We investigate to what extent this is the case. Figure~\ref{fig:PC2_outliers} shows PC2 values according to the percentile to which the examples belong. The weighting process indeed reduces the proportion of outliers. 

\begin{figure}[H]
  \centering
  \includegraphics[width=1\columnwidth]{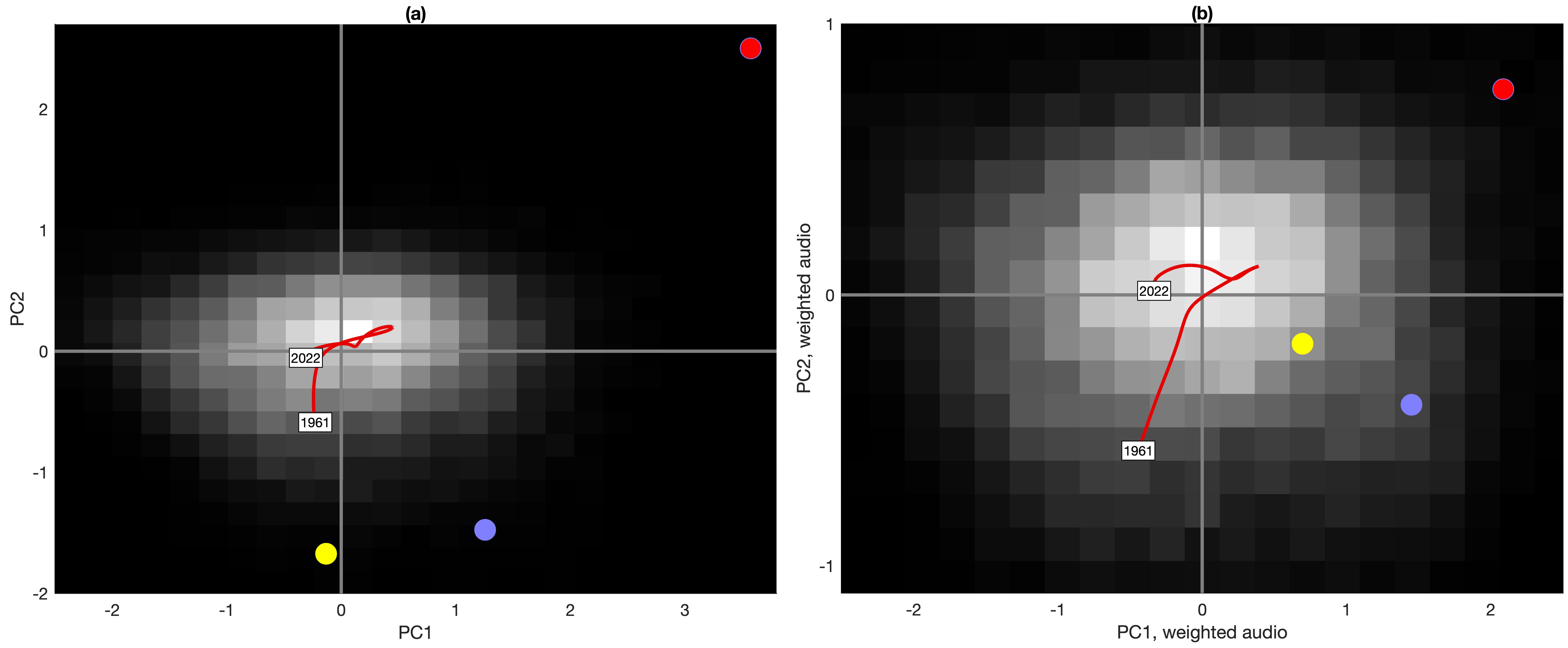}
  \caption{Positions of XXXTentacion's `Floor 555' (red), Danger Mouse's `No gold teeth' (blue), and Playboi Carti's `New tank' (yellow), in the representations from unweighted audio (a) and weighted audio (b).}
\label{fig:shiftfromphytoweight}
\end{figure}

Three examples of the phenomenon are shown in Figure~\ref{fig:shiftfromphytoweight}:

\begin{itemize}
    \item {\bf `Floor 555', by XXXTentacion \citep{xxxtentacionxx2018}}$\quad$ This track results in exceptionally high PC1 and PC2 values. There exist other tracks from the dataset that do not sound so different from `Floor 555' but that do not result in such extreme values. An example is `Surf Solar' by band Fuck Buttons \citep{fuckbuttons2009}, which also features noisy elements around the high-medium frequencies and a solid rhythm track with a prominent kick drum. A close examination of `Floor 555' shows that the extreme PC values result from low-frequency elements in the kick drum. Such elements cannot be heard at normal monitoring levels on most playback systems. Using weighted audio results in PC values that still correspond to inharmonic and noisy content but that are more commensurate to comparable-sounding tracks.

    \item {\bf`No gold teeth' by Danger Mouse \citep{dangermouse2022} and `New tank' by Playboi Carti \citep{playboicarti2021}}$\quad$ Both tracks have in common a loud, non-inharmonic bass, as well as higher frequency, more inharmonic elements. When weighted, the influence of the bass diminishes, and the relative inharmonicity (PC2) values are significantly increased.
\end{itemize}

Overall, we can observe a higher uniformity in terms of PC1 and PC2 when the corresponding content can be more easily perceived. In other words, music producers may be more conservative when they work with frequency bands for which the ear is more sensitive.


\section*{Acknowledgements}

We would like to thank Maarten Grachten\footnote{\url{https://scholar.google.es/citations?user=W95OZTYAAAAJ}} for his useful input. We are also grateful to Yann Mac\'e and Luc Leroy from the music production company, Hyper Music, for their helpful comments.

\section*{Disclosure statement}

The authors report there are no competing interests to declare.

\newpage

\bibliographystyle{apalike}
\bibliography{Harmonicity.bib}

\end{document}